\documentclass[prd,nofootinbib,english]{revtex4}
\usepackage{amsmath}
\usepackage{amsfonts,color}
\usepackage{amssymb,float}
\usepackage[utf8]{inputenc}
\usepackage{color}
\usepackage[unicode=true,pdfusetitle,
 bookmarks=true,bookmarksnumbered=true,bookmarksopen=true,bookmarksopenlevel=3,
 breaklinks=false,pdfborder={0 0 1},backref=false,colorlinks=true]
 {hyperref}
\usepackage{enumitem}
\usepackage{soul}

\usepackage{tikz}
\usetikzlibrary{shapes.geometric}
\usetikzlibrary{arrows.meta,arrows}

\usepackage[nice]{nicefrac}

\usepackage[outermarks]{titlesec}

\titleformat{\paragraph}
   {\normalfont\normalsize\centering}{\theparagraph .}{1em}{}
\titleformat{\subparagraph}
   {\normalfont\normalsize\bfseries}{\thesubparagraph}{1em}{}

\titlespacing*{\paragraph}    {0pt}{3.25ex plus 1ex minus .2ex}{1.5ex plus .2ex}
\titlespacing*{\subparagraph} {0pt}{3.25ex plus 1ex minus .2ex}{1.5ex plus .2ex}

\begin{document}

\title{Exact Spherically Symmetric Solutions in Modified Teleparallel gravity}

\author{Sebastian Bahamonde}\email{sbahamonde@ut.ee, sebastian.beltran.14@ucl.ac.uk}
\affiliation{Laboratory of Theoretical Physics, Institute of Physics, University of Tartu, W. Ostwaldi 1, 50411 Tartu, Estonia}
\affiliation{Department of Mathematics, University College London, Gower Street, London, WC1E 6BT, United Kingdom}
\author{Ugur Camci}\email{ugurcamci@gmail.com}
\affiliation{Siteler Mahallesi, 1307 Sokak, Ahmet Kartal Konutlari, A-1
Blok, No:7/2, 07070, Konyaalti, Antalya, Turkey}

\date{\today}

\begin{abstract}
\textbf{Abstract:} 
Finding spherically symmetric exact solutions in modified gravity is usually a difficult task. In this paper we use the Noether's symmetry approach for  a modified Teleparallel theory of gravity labelled as $f(T,B)$ gravity where $T$ is the scalar torsion and $B$ the boundary term. Using the Noether's theorem, we were able to find exact spherically symmetric solutions for different forms of the function $f(T,B)$ coming from the Noether's symmetries. 
\end{abstract}

\maketitle

\section{\uppercase{Introduction}}
General Relativity (GR) is a very successful theory but it has some shortcomings such as explaining and describing dark energy and dark matter~\cite{Copeland:2006wr,Bertone:2004pz}. Both the cosmological constant problem~\cite{Martin:2012bt} and the growing tension between the measure of the Hubble constant $H_0$ using direct and indirect measurements~\cite{Aghanim:2018eyx,Riess:2018uxu,Wong:2019kwg}, have increased the popularity of studying possible viable modifications of General Relativity and also the $\Lambda$CDM model. The recent discovery of the gravitational waves~\cite{Abbott:2016blz} and also the first image of the shadow of a supermassive black hole in the center of the galaxy M87~\cite{Akiyama:2019cqa}, have also increased the possibility of studying how gravity behaves under a strong regime. This effect might be also used to test GR and also to see possible effects that might arise in modified gravity. 

There are several modified theories of gravity, for some reviews, see~\cite{Clifton:2011jh,Heisenberg:2018vsk,Ishak:2018his,Nojiri:2017ncd}. The theory that we are interested in this paper is related to the so-called Teleparallel theories of gravity. In this framework, the torsion tensor is related to the gravitational field strength while the curvature tensor is zero~\cite{Aldrovandi:2013wha}. The Teleparallel equivalent of General Relativity is a specific theory in this framework that has the same equations as GR. One can then modify this theory and in the last years, different modified Teleparallel theories of gravity have been formulated and studied. One popular example is $f(T)$ gravity where $T$ is the scalar torsion~\cite{Cai:2015emx,Ferraro:2006jd}. There are several works about this theory in different contexts like cosmology~\cite{Bengochea:2008gz,Cai:2011tc,Bamba:2010wb,Dent:2011zz}, gravitational waves and astrophysics~\cite{Ferraro:2011ks,Nunes:2019bjq,Bamba:2013ooa,Cai:2018rzd,Nunes:2018evm,Farrugia:2018gyz,Ahmed:2016cuy,Boehmer:2011gw}. There are few works in this theory regarding vacuum spherically symmetric solutions. The first solutions were found using perturbation theory~\cite{Ruggiero:2015oka,DeBenedictis:2016aze}. In a recent paper, more general perturbed spherically symmetric solutions were found~\cite{Bahamonde:2019zea}. Still, there are few exact spherically symmetric solutions in modified Teleparallel gravity. Different authors have studied stars and wormholes in different Teleparallel theories~\cite{Bohmer:2011si,Bahamonde:2016jqq,Boehmer:2011gw,Horvat:2014xwa,Pace:2017aon}. In \cite{Ferraro:2011ks,Paliathanasis:2014iva}, the authors found some vacuum exact solutions in $f(T)$ gravity, however, they imposed $T=constant$, which is nothing but consider GR plus a cosmological constant. Still, there is a long route for finding new interesting exact solutions in modified teleparallel theory. 

In this paper, we study a generalise version of $f(T)$ gravity where now the boundary term $B$ that connects the scalar torsion $T$ with the Ricci scalar $\bar{R}$ is considered. This theory is called $f(T,B)$ gravity and resembles certain similitude with  $f(\bar{R})$ gravity~\cite{Bahamonde:2015zma}. In~\cite{Farrugia:2018gyz,Abedi:2017jqx} the authors found that there is one extra gravitational polarization mode in this theory, as in $f(\bar{R})$ gravity. It has been found that $f(T,B)$ cosmology can explain dark energy without evoking a cosmological constant and also that this theory can describe a transition from a matter dominated era to two different accelerated eras, one of which describes a de Sitter universe~\cite{Bahamonde:2016cul,Karpathopoulos:2017arc,Paliathanasis:2017flf}. 

Our main aim is to find exact spherically symmetric solutions in $f(T,B)$ gravity. One powerful tool that is useful for this is the so-called Noether's symmetry approach~\cite{Capozziello:1996bi}. One can use this method to obtain conserved quantities asking for the symmetries of the Lagrangian. The existence of some kinds of symmetry for the Euler-Lagrange equations of motion possessing a Lagrangian would
immediately be connected with the Noether symmetry. Noether symmetries are
capable of selecting suitable gravity theory and then to integrate dynamics by using the first integrals corresponding to
the Noether symmetries~\cite{Dialektopoulos:2018qoe}.
A consequent process with regard to the first integrals due to the Noether symmetries allows achieving exact solutions of the dynamical equations for the gravity theory. This theorem has been widely used in different contexts of modified gravity for obtaining exact solutions~\cite{Capozziello:2008ch,Atazadeh:2011aa,Capozziello:2012iea,Paliathanasis:2014iva,Bahamonde:2016jqq,Bahamonde:2017sdo,Bahamonde:2018ibz,Bahamonde:2016grb,Bahamonde:2018zcq,Dialektopoulos:2019mtr,Capozziello:2018gms,Capozziello:2016eaz,Paliathanasis:2011jq}. Our main aim is to use the Noether's symmetry approach to select the form of $f$ and then use the conserved charges to be able to solve the field equations and then obtain exact spherically symmetric solutions.

 In Sec.~\ref{tele} we give a brief introduction to Teleparallel gravity and more specifically to $f(T,B)$ gravity, and then, we find the corresponding spherically symmetric equations in this theory. Sec.~\ref{pointlike} is devoted to finding the point-like Lagrangian for the minisuperspace of a spherically symmetric spacetime. Sec.~\ref{noether} is the most important section where we use Noether's symmetry approach for our study case and find new exact solutions. Finally, we conclude our main results in Sec.~\ref{conclusions}. The notation of our paper is the following: Latin indices refer to the tangent space indices and Greek indices to the space-time indices. The metric signature used is $(+---)$.

\section{\uppercase{$f(T,B)$ gravity and Spherical symmetry}}
\label{tele}
Teleparallel gravity is an alternative framework of gravity where the manifold is globally flat and the torsion tensor is the responsible of the gravitational effects~\cite{Aldrovandi:2013wha,aldrovandi1995introduction,Aldrovandi:2003pa}. The fundamental variable in this framework is the tetrad fields $e^{a}{}_\mu$ that are the linear basis on the spacetime manifold, and at each point of the spacetime, gives us basis for vectors on the tangent space. The metric and its inverse can be reconstructed using the following relationships,
\begin{align}
  g_{\mu\nu} &= e^{a}{}_{\mu} e^{b}{}_{\nu} \eta_{ab} \,, \quad g^{\mu\nu} = E_{a}{}^{\mu} E_{b}{}^{\nu} \eta^{ab} \,,
\end{align} 
where $\eta_{ab}=\textrm{diag}(1,-1,-1,-1)$ is the Minwkoski metric and $E_{a}{}^{\mu}$ denotes the inverse of the tetrad. 

Within this framework, there exists an equivalent version of General Relativity (GR) containing the same Einstein's field equations labelled as the \textrm{Teleparallel equivalent of General Relativity} (TEGR). This theory is constructed from the action~\cite{Bahamonde:2015zma}
\begin{align}
\mathcal{S}_{\rm TEGR}= \frac{1}{2\kappa^2}\int d^4x e\,T\,, \label{taction}
\end{align}
where $\kappa^2=8\pi G $, $e=\textrm{det}(e^a{}_\mu)=\sqrt{-g}$ is the determinant of the tetrad and $T$ is the scalar torsion which is constructed from a contraction of the torsion tensor $T^{a}{}_{\mu\nu}=2 \left(\partial_{[\mu}e^a{}_{\nu]} + \omega^a{}_{b[\mu} e^b{}_{\nu]}\right)$ (where $\omega^a{}_{b\mu}$ is the spin connection), namely
\begin{align}
T=\frac{1}{4}T^{abc}T_{abc}+\frac{1}{2}T^{abc}T_{bac}-T^aT_a\,,
\end{align}
with $T_\mu=E_a{}^\nu T^a{}_{\nu\mu}$ being the torsion vector. The scalar torsion is directly connected to the curvature scalar computed with the Levi-Civita as~\cite{Bahamonde:2015zma,Bahamonde:2017wwk}
\begin{align}
 \bar{ R} = - T + \frac{2}{e}\partial_\mu (e T^\mu)=-T+B \,. \label{ricciT}
\end{align}
Here, $B$ is a boundary term, so that it is clear that the action \eqref{taction} differs only by a boundary term to the Einstein Hilbert action and then, TEGR give rise to the standard Einstein's field equations.

Even though TEGR has the same equations as GR, there are other teleparallel theories that have different field equations to GR and modified GR. One of the most promising theories in this framework is the so-called $f(T)$ theory where $T$ is replaced by an arbitrary function in the action \eqref{taction}~\cite{Cai:2015emx,Ferraro:2006jd}. A further generalisation of this theory also incorporates the boundary term by considering the action~\cite{Bahamonde:2015zma}
\begin{align}
\mathcal{S}_{f(T,B)}= \frac{1}{2\kappa^2}\int d^4x e\,f(T,B)\,, \label{taction2}
\end{align}
which contains both $f(T)$ gravity and also $f(\bar{R})$ gravity. This theory is a 4th order theory for the tetrad fields and since $T$ and $B$ are invariant under local Lorentz transformations, the above theory is also invariant under local Lorentz transformations. 

Let us now start with a spherically symmetric metric in spherical coordinates given by
\begin{align}
ds^2=a(r)^2 dt^2-b(r)^2 dr^{2} - M(r)^{2}d\Omega^{2}\,, \label{metric}
\end{align}
where $a(r), b(r)$ and $M(r)$ are functions of the radial coordinate $r$, and $d\Omega^2 = d\theta^2 + \sin^2 \theta d\varphi^2$. One  tetrad which reconstructs the metric via \eqref{metric} is the following
\begin{eqnarray}
e^{a}{}_{\mu}= \left(\begin{array}{cccc}
a(r) & 0 & 0 & 0 \\
0& b(r) \sin\theta\cos\varphi & M(r) \cos\theta\cos\varphi & -M(r) \sin\theta\sin\varphi\\
0& b(r) \sin\theta\sin\varphi & M(r) \cos\theta\sin\varphi & M(r) \sin\theta\cos\varphi\\
0& b(r) \cos\theta & -M(r)\sin\theta & 0
\end{array}\right)\label{tetrad2}\, .
\end{eqnarray}
The torsion tensor depends on both the tetrads and the spin connection, but the latter is a pure gauge term that is only related to the inertial effects. It has been shown that the above tetrad is compatible with a zero spin connection~\cite{Tamanini:2012hg,Hohmann:2019nat}. The torsion scalar and boundary terms for the above tetrad are given by
\begin{align}
T&= \frac{2}{M b^2}\left( b - M' \right) \left[ \frac{(b- M')}{M} - \frac{2 a'}{a} \right] \,, \label{T2} \\ B&= - \frac{4}{b M} \left(\frac{a'}{a} + \frac{ M'}{M} \right) + \frac{2}{b^2} \left[ \frac{a''}{a} - \frac{ a' b'}{ a b} +  \frac{2 M''}{M} + \frac{2 M'}{M} \left( \frac{2 a'}{a} - \frac{b'}{b}  + \frac{M'}{M} \right)  \right]\,.\label{B2}
\end{align}
Here primes denote differentiation with respect to the radial coordinate. Clearly, this tetrad also gives the correct Minkowski limit ($a=b=1$) for the above scalars $T=B=0$. It is important to remark that it is equivalent to consider the tetrad~\eqref{tetrad2} with a zero spin connection to consider a diagonal tetrad with a non-zero spin connection as in~\cite{Krssak:2015oua}.

As expected, by subtracting the above equations, one recovers the standard scalar curvature computed with the Levi-Civita connection,
\begin{align}
\bar{R}& = -T + B = -\frac{2}{M^2} + \frac{2}{b^2} \left[ \frac{a''}{a} + \frac{ 2 M''}{M} - \frac{a' b' }{ a b}  + \frac{2 a' M' }{a M} -\frac{2 b' M'}{b M} + \frac{ M'^2}{M^2}  \right]\,.
\end{align}
By varying the $f(T,B)$ action with respect to the tetrads, one gets the following field equations in vacuum~\cite{Bahamonde:2015zma}
\begin{multline}
  2e\delta_{\nu}^{\lambda}\Box f_{B}-2e\nabla^{\lambda}\nabla_{\nu}f_{B}+
  e B f_{B}\delta_{\nu}^{\lambda} + 
  4e\Big[(\partial_{\mu}f_{B})+(\partial_{\mu}f_{T})\Big]S_{\nu}{}^{\mu\lambda}
  \\
  +4e^{a}_{\nu}\partial_{\mu}(e S_{a}{}^{\mu\lambda})f_{T} - 
  4 e f_{T}T^{\sigma}{}_{\mu \nu}S_{\sigma}{}^{\lambda\mu} - 
  e f \delta_{\nu}^{\lambda} = 0 \,,
  \label{fieldeq}
\end{multline}
where $f_T=\partial f/\partial T$, $f_B=\partial f/\partial B$, $\Box=\nabla^\mu\nabla_\mu$ and $ S^{abc} = \frac{1}{4}(T^{abc}-T^{bac}-T^{cab})+\frac{1}{2}(\eta^{ac}T^b-\eta^{ab}T^c)$ is the superpotential. By replacing the tetrad~\eqref{tetrad2} into the above field equations, we find that the vacuum spherically symmetric field equations in $f(T,B)$ gravity are given by
\begin{eqnarray}
f - B f_B - \frac{4}{b^2} f_T \left[ \frac{M''}{M} + \frac{M'^2}{M^2} + \frac{M'}{M} \left( \frac{a'}{a} - \frac{ b'}{b} \right) - \frac{b}{M} \left( \frac{a'}{a} + \frac{b'}{b} \right) \right] + \frac{4}{M b^2} f_T' \left( b - M' \right)&& \nonumber \\ \quad + \frac{2}{b^2} f'_B \left( \frac{2 b}{M}  - \frac{b'}{b} \right) + \frac{2}{b^2} f_B''&=&0  \, ,  \label{feq-1} \\ f - B f_B -  T f_T - \frac{2}{b^2} f_T \left( \frac{M'^2}{M^2} + \frac{2 a' M'}{a M} - \frac{b^2}{M^2} \right) + \frac{2}{b^2} f_B' \left(  \frac{a'}{a} + \frac{2 M'}{M} \right)  &=&0\,, \label{feq-2}  \\
f - B f_B - \frac{2}{b^2} f_T \left[ \frac{a''}{a} + \frac{M''}{M} + \frac{M'^2}{M^2} + \frac{M'}{M} \left( \frac{3 a'}{a} - \frac{b'}{b} \right) - \frac{a' b'}{a b}  - \frac{2 b}{M} \left( \frac{a'}{a} + \frac{M'}{M} \right)  + \frac{b^2}{M^2} \right]&&  \nonumber\\
 \quad - \frac{2}{b^2} f_T' \left( \frac{a'}{a} + \frac{M'}{M} - \frac{b}{M} \right)  + \frac{2}{b^2} f_B' \left( \frac{b}{M}  - \frac{b'}{b} \right) + \frac{2}{b^2} f_B''&=&0  \,.  \label{feq-3}
\end{eqnarray}
Our aim is to use the Noether's symmetry approach to get some exact solutions for the above system of partial differential equations. In the next section, we will find the point-like Lagrangian which is an ingredient needed for this purpose.

\section{\uppercase{The point-like Lagrangian}}\label{pointlike}

In order to find the point-like Lagrangian in terms of the configuration space variables $a, b, M, T$ and $B$, we use the Lagrange multiplier method, and write the canonical action of $f(T,B)$ gravity as follows
\begin{eqnarray}
\mathcal{S}_{f(T,B)} &=& \int{ dr \mathcal{L}_{f(T,B)} (r, a, b, M, T, B, a', b', M', T', B') }\,,  \label{action1} \\ & = & \int{ dr  \sqrt{|g|} \left\{  f(T,B) - \lambda_1 \left[ T - \bar{T} \right] - \lambda_2 \left[ B - \bar{B} \right] \right\} }\,, \label{action2}
\end{eqnarray}
where $\lambda_1$ and $\lambda_2$ are the Lagrangian multipliers,  $\bar{T}$ and $\bar{B}$
are, respectively, the torsion scalar and the boundary term given in \eqref{T2} and \eqref{B2}. The Lagrange multipliers $\lambda_1$ and $\lambda_2$ are obtained by varying the action \eqref{action2} with respect to $T$ and $B$, respectively. By doing this, one gets $\lambda_1 = f_T$ and $\lambda_2 = f_B$. Then the $f(T,B)$ gravity action for the tetrad~\eqref{tetrad2} becomes
\begin{eqnarray}
\mathcal{S}_{f(T,B)} & = &  \int dr  a b M^2 \Big\{  f(T,B) - f_T \left[ T - \frac{2 }{a b^2 M^2} \left( b - M' \right) \left(-2 M a' - a M' + a b \right) \right]  \nonumber \\ & & - f_B \Big[ B +  \frac{4}{b M} \left(\frac{a'}{a} + \frac{ M'}{M}\right) - \frac{2}{b^2} \left( \frac{a''}{a} - \frac{ a' b'}{ a b} +  \frac{2 M''}{M} + \frac{4 a' M'}{a M} - \frac{2 b' M'}{b M}  + \frac{2 M'^2}{M^2} \right)\Big]  \Big\}\, . \qquad   \label{action3}
\end{eqnarray}
Therefore, after discarding divergence terms by integrating by parts, the point-like Lagrangian can be expressed as follows
\begin{eqnarray}
\mathcal{L}_{f(T,B)} &=& - 4 M ( f_T + f_B ) a' - 4 a (f_T + f_B) M' + \frac{2}{b} f_T \left( 2 M a' M' +  a M'^2 \right)   \nonumber\\ && - \frac{ 2 M }{b} f_B' \left( M a'+ 2 a M' \right) - a b  \left[  M^2 ( T f_T + B f_B - f ) - 2 f_T \right]   \, ,
\label{L}
\end{eqnarray}
where $f'_B = f_{BT} T' + f_{BB} B'$.
Because of the absence of the generalized velocity $b'$ in the point-like Lagrangian \eqref{L}, the Hessian determinant of this Lagrangian vanishes, as it should be.  By varying the point-like Lagrangian density \eqref{L} with respect to $a, b$ and $M$, we find respectively
\begin{eqnarray}
& & 2 f_T \left[ \frac{2 M''}{M} + \frac{M'^2}{M^2} - \frac{2 b' M'}{b M} - \frac{b^2}{M^2} \right] + \frac{4 f'_T}{M}  \left( M' - b \right) + 2 f'_B \left( \frac{b'}{b} - \frac{2 b}{M} \right) - 2 f''_B - b^2 \left( f- T f_T - B f_B \right)  = 0, \qquad   \label{var-a}  \\ & & 2 f_T \left[  \frac{ M'}{M} \left( \frac{2 a'}{a} +  \frac{M'}{M} \right) - \frac{b^2}{M^2} \right] - 2 f'_B \left(  \frac{a'}{a} +  \frac{2 M'}{M} \right) - b^2 \left( f - T f_T - B f_B \right)  = 0\,, \label{var-b}
\end{eqnarray}
and
\begin{eqnarray}
& & 2 f_T \left[ \frac{a''}{a} +\frac{ M''}{M} + \frac{a' M'}{a M} - \frac{b' M'}{b M} - \frac{a' b'}{a b} \right] + 2 f'_T \left( \frac{a'}{a} + \frac{M'}{M} - \frac{b}{M} \right) + 2 f'_B \left( \frac{b'}{b} - \frac{b}{M} \right) \nonumber \\ & & \qquad -  2 f''_B - b^2 \left( f - T f_T - B f_B \right)  = 0\,.  \label{var-M}
\end{eqnarray}
Then, variations with respect to $T$ and $B$ yield
\begin{eqnarray}
& & \left[ T - \bar{T} \right] f_{TT} + \left[ B - \bar{B} \right] f_{BT} = 0,  \label{var-T}  \\ & & \left[ T - \bar{T} \right] f_{TB} + \left[ B - \bar{B} \right] f_{BB} = 0\,, \label{var-B}
\end{eqnarray}
which shows a symmetry in the variables $T$ and $B$. As is usually done in cosmology, one can also rewrite the field equations in a different form by assuming that the new contributions coming from modified $f(T,B)$ are related to a fluid with a pressure and energy density. Assuming $f_T \neq 0$, one can rewrite Eqs. \eqref{var-a}, \eqref{var-b} and \eqref{var-M} as
\begin{eqnarray}
\rho_{_{TB}}&= & \frac{2}{b^2} \left( \frac{2 b' M'}{b M} - \frac{2 M''}{M} - \frac{M'^2}{M^2}  + \frac{b^2}{M^2} \right)\,,  \label{var-a-2}  \\ p_{_{TB}}&= & \frac{2}{b^2} \left(  \frac{ 2 a' M'}{a M} + \frac{M'^2}{M^2}  - \frac{b^2}{M^2} \right) \,,  \label{var-b-2}
\end{eqnarray}
and
\begin{eqnarray}
& & f_T \left( \frac{a''}{a} - \frac{M''}{M} - \frac{M'^2}{M^2} + \frac{a' M'}{a M} + \frac{b' M'}{b M} - \frac{a' b'}{a b}  + \frac{b^2}{M^2} \right) + f'_T  \left( \frac{a'}{a} - \frac{M'}{M} + \frac{b}{M} \right) + \frac{b}{M} f'_B = 0\,,  \qquad  \label{var-M-2}
\end{eqnarray}
where $\rho_{_{TB}}$ and  $p_{_{TB}}$ are defined as
\begin{eqnarray}
& & \rho_{_{TB}} \equiv \frac{1}{f_T} \left[ T f_T + B f_B - f   + \frac{4}{b^2} f'_T  \left( \frac{M'}{M} - \frac{b}{M} \right) + \frac{2}{b^2} f'_B \left( \frac{b'}{b} - \frac{2 b}{M} \right) - \frac{2}{b^2} f''_B \right]\,,    \label{def-rho}  \\ & & p_{_{TB}} \equiv \frac{1}{f_T} \left[ f - T f_T - B f_B  - \frac{2}{b^2} f'_B \left(  \frac{a'}{a} +  \frac{2 M'}{M} \right) \right]\,, \label{def-p}
\end{eqnarray}
which are the torsion and the boundary term contributions to energy density and pressure. Clearly, if we neglect all contributions from $f(T,B)$, i.e., $\rho_{TB}=p_{TB}=0$, from Eqs.~\eqref{var-a-2}-\eqref{var-b-2} we find that $a(r)=\sqrt{1-2M/r}=1/b(r)$ which is the Schwarzschild solution in General Relativity. Based on the field equations \eqref{var-a-2}-\eqref{var-M-2}, we can derive the conservation equation. Here we follow the well known derivation of the conservation equation in General Relativity. By differentiating \eqref{var-b-2}, one gets
\begin{eqnarray}
& & p'_{_{TB}} = \frac{b^2}{4} \left\{ \frac{M'}{M} \left( \frac{a''}{a} + \frac{M''}{M} - \frac{a' M'}{a M} -  \frac{b' M'}{b M}  - \frac{M'^2}{M^2} + \frac{b^2}{M^2} \right) +  \frac{a'}{a} \left( \frac{M''}{M} - \frac{a' M'}{a M} - \frac{2 b' M'}{b M} \right) \right\}\,,   \label{drv-p}
\end{eqnarray}
Then, the combination $\rho_{_{TB}} + p_{_{TB}}$ can be obtained from the field equations \eqref{var-a-2}-\eqref{var-b-2}, namely
\begin{eqnarray}
& & \rho_{_{TB}} + p_{_{TB}} = -\frac{4}{b^2} \left( \frac{M''}{M} - \frac{a' M'}{a M} - \frac{b' M'}{b M} \right) \, ,  \label{rho+p}
\end{eqnarray}
which yields from the equation \eqref{drv-p} that
\begin{eqnarray}
& & p'_{_{TB}} + \left( \frac{a'}{a} + \frac{M'}{M} \right) \left( \rho_{_{TB}} + p_{_{TB}} \right) = \frac{4}{b^2} \frac{M'}{M} \left( \frac{a''}{a} - \frac{a' b'}{a b} - \frac{M'^2}{M^2} + \frac{b^2}{M^2} \right) \, .  \label{conv-eq1}
\end{eqnarray}
Thus, the field equation \eqref{var-M-2} allows us to write the latter equation as
\begin{eqnarray}
& & p'_{_{TB}} + \left( \frac{a'}{a} + \frac{M'}{M} \right) \left( \rho_{_{TB}} + p_{_{TB}} \right) = - \frac{4}{b^2 f_T} \frac{M'}{M} \left[ f'_{T} \frac{a'}{a}  + f'_B \left( \frac{3 a' }{2 a} + \frac{ b'}{2 b} + \frac{3 M'}{M} \right) - \frac{1}{2} f''_{B} \right] \, .  \label{conv-eq2}
\end{eqnarray}
Remembering here that the ordinary matter is not included in the field equations, i.e., $\rho = p_r = p_t = 0$. Therefore, we are not expecting that the conservation equation \eqref{conv-eq1} or \eqref{conv-eq2} due to the dark sector of the universe has the usual form. 

The energy function associated with the Lagrangian $\mathcal{L}$, which is known as the Hamiltonian of the system, is defined by
\begin{equation}
E_{\mathcal{L}} = q'^i \frac{\partial \mathcal{L}}{\partial q'^i} -\mathcal{L}\,, \label{energy}
\end{equation}
where $q^i = \left\{ a, b, M, T, B \right\}$ are generalized coordinates for the Lagrangian density \eqref{L} of the $f(T,B)$ theory of gravity. Then, computations show that   $E_{\mathcal{L}}$ vanishes, because of Eq.~\eqref{var-b} due to the variation with respect to $b$. The equation $E_{\mathcal{L}} = 0$ can be explicitly solved in terms of $b$ as a function of the remaining generalized coordinates, yielding
\begin{eqnarray}
& & b(r)^2 = \frac{  f_T M M' \left(  \frac{2 a'}{a} + \frac{M'}{M} \right) - M^2 f'_B \left( \frac{ a'}{a} + 2 \frac{M'}{M} \right) }{ \frac{M^2}{2} ( f - T f_T - B f_B ) +  f_T  }\,. \label{e-b1}
\end{eqnarray}
The next section will be devoted to using the Noether's symmetry approach to get exact solutions for different forms of $f(T,B)$.

\section{\uppercase{Noether symmetry approach and exact solutions}}\label{noether}
In this study, we consider the Noether symmetry generator which has the form
\begin{equation}
{\bf X} = \xi (r, q^k) \frac{\partial}{\partial r} + \eta^i (r, q^k) \frac{\partial}{\partial q^i}\,, \label{ngs-gen}
\end{equation}
where $q^i = \left\{ a, b, M, T, B \right\}, \, i \in \left\{ 1,2,3,4,5 \right\}$, are the generalized coordinates in the $5$-dimensional configuration space ${\cal Q }\equiv \{ q^i, i=1, \ldots, 5 \} $ of the Lagrangian, whose tangent space is ${\cal TQ }\equiv \{q^i,q'^i\}$.  The existence of a Noether symmetry implies the existence of a vector field ${\bf X}$ given in (\ref{ngs-gen}) if the Lagrangian $ \mathcal{L}(r, q^i, q'^i )$ satisfies the condition
\begin{equation}
{\bf X}^{[1]} \mathcal{L} + \mathcal{L} ( D_r \xi) = D_r K\, , \label{ngs-eq}
\end{equation}
where ${\bf X}^{[1]}$ is the first prolongation of the generator (\ref{ngs-gen}) in such a form
\begin{equation}
{\bf X}^{[1]} = {\bf X}  + \left( D_r \eta^i - q'^i D_r \xi \right)  \frac{\partial}{\partial q'^i}\,.
\end{equation}
Here, $K(r, q^i)$ is a gauge function, and $D_r$ is the total derivative operator with respect to $r$, defined by $D_r =\partial / \partial r + q'^i \partial / \partial q^i$. The significance of a Noether symmetry comes from the first integral of motion. If ${\bf X}$ is the Noether symmetry generator corresponding to the Lagrangian $\mathcal{L}(r, q^i, q'^i)$, then a conserved quantity associated with the generator ${\bf X}$ is
\begin{equation}
I = - \xi E_{\mathcal{L}} + \eta^i \frac{\partial \mathcal{L}}{\partial q'^i} - K\, , \label{con-law}
\end{equation}
where $I$ is a constant of motion.

The Noether symmetry condition \eqref{ngs-eq} for the point-like Lagrangian \eqref{L} gives an over-determined system of 34 partial differential equations. They are shown in the appendix for completeness (see Appendix~\ref{appendix}). These differential equations will fix the form of the vector ${\bf X}$ as well as the form of $f(T, B)$. Depending on the function $f(T,B)$, it is possible to study several different branches, such that \begin{itemize}
    \item Case 1: $f(T,B) = F_0 (T) + F_1 (B)$ ($f_{TB} = 0$).
    \item Case 2:  $f(T,B)= T\, h_1 (B) + h_2 (B)$ ($f_{TT}=0$).
    \item Case 3: $f(T,B)= g_1 (T) + B g_2 (T)$ ($f_{BB} = 0$).
\item Case 4: $f_{TB}\neq 0$ (general case).
\end{itemize}
Then, a Noether symmetry exists if at least one of the coefficients $\xi, \eta^1, \eta^2, \eta^3, \eta^4$ and $\eta^5$ is different from zero. We summarize them in each of the mentioned cases like the following. For any case, we have the Noether symmetry
\begin{equation}
{\bf X}_0 = \alpha (r) \partial_r - 2 \alpha'(r) \partial_b\,,
\end{equation}
where $\alpha (r)$ is an arbitrary function of $r$. Using \eqref{con-law}, this Noether symmetry yields the energy conservation relation $E_{\mathcal{L}} = 0$ if $\alpha (r) \neq 0$. In the following, we will split the study for all the cases mentioned above. 

\subsubsection{Case 1: $f(T,B) = F_0 (T) + F_1 (B)$ ($f_{TB} = 0$)}
In this case, we assume the condition $f_{TB} = 0$ giving \begin{equation}
     f(T,B) = F_0 (T) + F_1 (B)\,.
\end{equation}
By using this condition into the Noether's equations \eqref{noether-eqs}, we find that the only possible solutions for $f$ which satisfy the Noether's conditions are given by:
\begin{equation}
     F_0 (T) = f_0 T^n\,,\quad F_1(B) = f_1 B^m\,,
\end{equation}
where $n$ and $m$ are real numbers. Depending on the parameters, we can find different Noether's vectors and different solutions.

    \paragraph{Subcase: $f(T,B)=f_0 T + f_1 B$ - General Relativity ($n=m =1$)}

For this subcase we find {\it four} Noether symmetries such as ${\bf X}_0$, and
\begin{eqnarray}
& & {\bf X}_1 = - a \partial_a + b \partial_b + M \partial_M\,, \label{X1}  \\& &  {\bf X}_2 = \frac{1}{a M} \left( \partial_a - \frac{b}{a} \partial_b \right) +  \partial_M  \quad {\rm with} \quad K = - \frac{4}{a} (f_0 + f_1)\,,  \label{X2} \\& & {\bf X}_3 = \frac{1}{2 M} \left( - a \partial_a + b \partial_b \right) + \partial_M  \quad {\rm with} \quad K = -2 (f_0 + f_1) a\,.  \label{X3}
\end{eqnarray}
Then, the corresponding Noether constants become
\begin{eqnarray}
& & I_1 = 4 f_0 M^2 \frac{a'}{b},  \quad  I_2 = 4 f_0 \frac{M'}{a b} , \quad I_3 = \frac{2 f_0 M}{a b} \left(  \frac{2 a'}{a} + \frac{M'}{M} \right)\,,   \label{NC123}
\end{eqnarray}
and $b^2 = M^2 \left(  \frac{2 a' M'}{a M} + \frac{M'^2}{M^2} \right)$ due to $E_{\mathcal{L}} = 0$. After solving the above first integrals, we obtain that
\begin{eqnarray}
& & a^2 = \frac{2 I_3}{I_2} \left( 1 - \frac{ \ell }{M}  \right),  \quad  b^2 = \frac{ M'^2 }{ 1 - \frac{\ell }{M}} ,  \quad I_2 I_3 = 8 f_0^2 \, ,   \label{sol-1}
\end{eqnarray}
where $\ell = I_1 / I_3$. This solution becomes the Schwarzschild solution for $M(r) = r$.

    \paragraph{Subcase: $f(T,B)=f_0 T^n$ - power-law $f(T)$ gravity ($n$ arbitrary and $m= 1$)}
In the case where $n$ is arbitrary and $m= 1$, the boundary term does not affect the field equations since $B$ appears linearly in the action, i.e., $f(T,B)=f_0 T^n + f_1 B$. Thus, this subcase corresponds to a power-law $f(T)$ gravity. For these theories, the field equations \eqref{var-a-2} and  \eqref{var-b-2} have the same form, and the field equation \eqref{var-M-2} becomes
\begin{eqnarray}
& &  \frac{a''}{a} - \frac{a' b'}{a b} - \frac{M'^2}{M^2} + \frac{b^2}{M^2} +  (n-1)\frac{a' T'}{a T}  = 0\,,   \label{feq-M-i}
\end{eqnarray}
together with
\begin{eqnarray}
& & \rho_{_{TB}} = (n-1) \left[ \frac{T}{n}  + \frac{4}{b^2} \frac{T'}{T}  \left( \frac{M'}{M} - \frac{b}{M} \right) \right], \qquad p_{_{TB}} = \frac{(1-n)}{n} T \, ,   \label{def-rho-p-i}  
\end{eqnarray}
which yields
\begin{eqnarray}
& &  p'_{_{TB}} + \frac{b^2 M T}{n (M' -b)} \left( \rho_{_{TB}} +  p_{_{TB}} \right) = 0 \, , \qquad b \neq M' \,.  \label{cnv-eq-i1}
\end{eqnarray}
Furthermore, Eq.~\eqref{conv-eq2} for this case becomes
\begin{eqnarray}
& & p'_{_{TB}} + \left( \frac{a'}{a} + \frac{M'}{M}  \right) \left( \rho_{_{TB}} + p_{_{TB}} \right) = \frac{4 (1-n)}{b^2} \frac{a' M' T'}{a M T}  \, .  \label{cnv-eq-i2}
\end{eqnarray}
Thus, by combining Eqs.~\eqref{cnv-eq-i1} and \eqref{cnv-eq-i2}, we obtain 
\begin{eqnarray}
& & \left[ \frac{4(n-1)}{b^2} \frac{a' M'}{a M} +  p_{_{TB}} \right] p'_{_{TB}} + \left( \frac{a'}{a} + \frac{M'}{M}  \right) p_{_{TB}} \left( \rho_{_{TB}} + p_{_{TB}} \right) = 0  \, .  \label{cnv-eq-i3}
\end{eqnarray}
If we assume a barotropic equation of state $p_{_{TB}} = w_{_{TB}} \rho_{_{TB}}$, then  Eq.~\eqref{cnv-eq-i3} reduces to the Abel differential equation of second kind for $ \rho_{_{TB}}$.

The Noether symmetries for this case are ${\bf X}_0$ and
\begin{eqnarray}
& & {\bf X}_1 = (2n-3) a \partial_a +  b \partial_b +  M \partial_M - 2 T \partial_T \quad {\rm with} \quad K = - 8 f_1 (n-1) M a \,. \label{X1-3}
\end{eqnarray}
The corresponding Noether first integrals give 
\begin{eqnarray}
& & b^2 = \frac{  M^2 \left( \frac{2 a' M'}{a M} + \frac{M'^2}{M^2} \right) }{ 1 + \frac{(1-n)}{2 n} M^2 T }, \qquad  I_1 =  \frac{4 f_0 a }{b} M^2 T^{n-1} \left[  \frac{a'}{a} +  2 (n-1) \left( \frac{M'}{M} - \frac{b}{M} \right) \right]\, .   \label{NC12-3}
\end{eqnarray}
There are two equations in \eqref{NC12-3}, but there are four unknowns $a, b, M$ and $T$. Using the definition of $T$ (see Eq.~\eqref{T2}) in the first equation of \eqref{NC12-3} yields in
\begin{eqnarray}
& & (1 - 2n) \left( \frac{2 a' M'}{a M} + \frac{M'^2}{M^2} \right) + 2 (n-1) \frac{b}{M} \left(  \frac{a'}{a} + \frac{M'}{M} \right) + \frac{b^2}{M^2} = 0\,.   \label{ceq-i-1}
\end{eqnarray}
Then, we can consider the latter equation to generate solutions by choosing the form of $M$, and this choice reduces the number of unknowns to two, $a$ and $b$. Without losing generality, we can choose $M(r)=r$. Then the Eq. \eqref{ceq-i-1} becomes
\begin{eqnarray}
& & 2 \left[ 1 - 2n + (n-1)b \right] \frac{a'}{a} + \frac{ \left[ 1 - 2n + 2 (n-1)b + b^2 \right]}{r} = 0\,.   \label{ceq-i-M-r1}
\end{eqnarray}
Also, the second equation in \eqref{NC12-3} has the form
\begin{eqnarray}
& &  \frac{a'}{a}  + 2 (n-1) \frac{(1-b)}{r} - \frac{ I_1 b \, T^{1-n}}{ 4 f_0 n  r^2 a } = 0\,,    \label{ceq-i-M-r2}
\end{eqnarray}
and $T$ becomes
\begin{equation}
T = \frac{2 (b-1)}{r^2 b^2} \left( b -1 - 2 r \frac{a'}{a} \right).  \label{T-i-2}
\end{equation}
Then, we immediately found  the following solution
\begin{eqnarray}
& & a(r) = \sqrt{ 1- \frac{k}{r}}, \quad b(r) = \frac{1}{\sqrt{ 1- \frac{k}{r}}},  \quad T(r) = -\frac{4}{r^2} \left(  \frac{1 - \frac{k}{2r}}{\sqrt{1 - \frac{k}{r}}} -1 \right), \quad  \rho_{_{TB}} = 0, \,\, p_{_{TB}} = 0, \,\, I_1 = 2 k f_0, \label{sol-i-M-r1}
\end{eqnarray}
for $n=1$, which is obviously the Schwarzschild solution with $k=2M$ and $f(T,B) = f_0 T + f_1 B$ theory (GR case).  For $n \neq \frac{1}{2}, \frac{5}{6},\frac{5}{4}, \frac{3}{2}$ and $M(r)= r$,  the following exact solution appears when $I_1 = 0$,
\begin{eqnarray}
& & a(r) = \left( \frac{r}{r_0} \right)^{ \frac{4 n (n-1)(2n -3)}{4 n^2 -8 n +5}}, \quad b(r) = k,  \qquad T(r) = -\frac{T_0}{r^2},\label{sol-i-M-r1} \\ & & \rho_{_{TB}} = \frac{\rho_0}{r^2}, \quad p_{_{TB}} = \frac{p_0}{r^2}, \qquad w_{_{TB}} = \frac{ (2n -3)( 4n -5)}{6n -5} \, , \label{sol-i-M-r2}
\end{eqnarray}
where $r_0$ is an integration constant of dimension length, $k$ is a dimensionless constant, $T_0, \rho_0$ and $p_0$ are as follows
\begin{eqnarray}
& & k = \frac{ (2n-1)(4n -5)}{4 n^2 - 8n +5}, \, T_0 = \frac{ 8 n^2 (2n -3)^2 }{(4n -5) (2n -1)^2}, \,  \rho_0 = \frac{ 8n (n-1) (2n -3)( 6n -5)}{(2n -1)^2 (4n -5)^2}, \,\,  p_0 = \frac{ 8n (n-1) (2n -3)^2 }{(4n -5) (2n -1)^2} \,. \qquad
\end{eqnarray}
The above solution is similar to the one found in \cite{Boehmer:2011gw}, in which the solution is given by the Eq.~(4.20). However, our solution is most general in the sense that $T(r) \neq 0$ here. Inspiring by some values for the equation of state parameter $w(= p / \rho)$ in the ordinary matter such as $w=1$ for stiff fluid and $w= \frac{1}{3}$ for the radiation, one can make a connection between $w_{_{TB}}$ and the dark sector of the universe. It is interesting to note that the equation of state parameter $w_{_{TB}}$ is a constant, and the power $n$ of $T$ has real values for $w_{_{TB}} > -\frac{13}{9} + \frac{4}{9} \sqrt{10}$ and $w_{_{TB}} < -\frac{13}{9} - \frac{4}{9} \sqrt{10}$, which means that $w_{_{TB}} = -1, -\frac{1}{3}$ give rise to imaginary values of $n$. Also, it is not possible to get $w_{_{TB}} = 0$ due to the restrictions on $n$. We point out that $w_{_{TB}}=1$ (\emph{the dark stiff fluid}) if $n=\frac{5}{2}$, and $w_{_{TB}}= \frac{1}{3}$ (\emph{the dark radiation}) if $n= \frac{3}{2} \pm \frac{1}{\sqrt{6}}$.

Furthermore, by assuming $I_1\neq0$, we can directly find the form of $b(r)$ by solving Eq.~\eqref{ceq-i-M-r1}, which gives us \begin{equation}
    b(r)=-\frac{(n-1) \left(r a'+a\right)}{a}\pm \frac{1}{a}\sqrt{2 n^2 r a a'+(n-1)^2 r^2 a'{}^2+n^2 a^2}\,.
\end{equation}
Then, the explicit form of $b(r)$ depends on $a(r)$ that in principle, one can find from Eq.~\eqref{ceq-i-M-r2}, giving us the following differential equation,
\begin{eqnarray}
&&2 f_0 (n-1) n r a \left((n-1) r a'+n a-\sqrt{2 n^2 r a a'+(n-1)^2 r^2 a'^2+n^2 a^2}\right)+f_0 n r^2 a a'\nonumber\\
&&+\frac{ 4^{-n} n I_1}{(2 n-1)(a (r-2 n r)^2 \left(2 r a'+a\right)) r^2} \left(n r a'+n a-\sqrt{2 n^2 r a a'+(n-1)^2 r^2 a'^2+n^2 a^2}\right)\times \nonumber \\ 
&&\Big[-n \Big(a \left(\sqrt{2 n^2 r a a'+(n-1)^2 r^2 a'^2+n^2 a^2}+2 n r a'\right)\nonumber\\
&&-r a' \left(\sqrt{2 n^2 r a a'+(n-1)^2 r^2 a'^2+n^2 a^2}+(n-1) r a'\right)+n a^2\Big)\Big]^{-n}=0\,. \label{a-diff-case-1a}
\end{eqnarray}
Now, we have one differential equation for one variable $a(r)$. The problem of finding solutions, is that this differential equation for $a(r)$ cannot be solved easily.

    \paragraph{Subcase: $f(T,B)=f_1 B^m$ -  power-law $f(B)$ gravity ($f_T = 0$)}
For the subcase $f_T = 0$, the field equations \eqref{var-a}--\eqref{var-M} are reduced to
\begin{eqnarray}
& & (m-1) \left[ \left( \frac{a'}{a} + \frac{2 M'}{M} \right) \frac{B'}{B} - \frac{b^2}{2m} B \right] = 0\,,  \label{feq-ia-1} \\ & & (m-1) \left[ \frac{B''}{B} + (m-2) \frac{B'^2}{B^2} - \left( \frac{b'}{b} - \frac{2 b}{M} \right) \frac{B'}{B} - \frac{b^2}{2m} B \right]= 0\,, \label{feq-ia-2} \\ & & (m-1) \left[ \frac{B''}{B} + (m-2) \frac{B'^2}{B^2} - \frac{1}{2} \left( \frac{b'}{b} - \frac{b}{M} \right) \frac{B'}{B} - \frac{b^2}{2m} B \right] = 0 \,. \label{feq-ia-3}
\end{eqnarray}
It is easy to see from the above equations that the boundary term $B$ does not contribute to the field equations for $m=1$. Then, using \eqref{feq-ia-1} and taking $m \neq 1$, we can find from the field equations \eqref{feq-ia-2} and \eqref{feq-ia-3} that
\begin{eqnarray}
& & \left( \frac{b'}{b} - \frac{3 b}{M} \right) B' = 0\,, \label{feq-ia-2-2} \\ & & \frac{B''}{B} + (m-2) \frac{B'^2}{B^2} - \left( \frac{a'}{a} + \frac{2 M'}{M} + \frac{b}{M} \right) \frac{B'}{B}  = 0 \,. \label{feq-ia-3-3}
\end{eqnarray}
Thus, Eq.~\eqref{feq-ia-2-2} gives $b(r)$ after choosing $M$ for $B' \neq 0$. For $m \neq 0, 1, 1/2$, we find the following {\it five} Noether symmetries such that ${\bf X}_0$ and
\begin{eqnarray}
& &  {\bf X}_1 = (2 m -3) \frac{a}{2} \partial_a - \frac{b}{2} \partial_b - \frac{M}{2} \partial_M  + B \partial_B\,,  \label{NS-ib-1} \\ & & {\bf X}_2 = -\frac{1}{a M^3} \left( \frac{1}{M} \partial_a + \frac{b}{a M} \partial_b - \frac{1}{a} \partial_M \right)\,, \label{NS-ib-2}   \\ & & {\bf X}_3 =  m B^{1-2m} \left( 2 a \partial_a -  b \partial_b -  M \partial_M + B \partial_B \right) \quad {\rm with} \quad K= - 4 f_1 m (2m -1) a M B^{-m} \,, \label{NS-ib-3}  \\ & & {\bf X}_4 = \frac{1}{a M^3} \left[ \frac{\mu'}{b}  - \frac{\mu}{M} \right] \partial_a - \frac{\mu  b}{a^2 M^4}  \partial_b + \frac{1}{a^2 M^2} \left[ \frac{\mu }{M} - \frac{ \mu' }{2 b} \right] \partial_M \quad  {\rm with} \,\, K = - 2 f_1 \frac{m}{a b M^2} B^{m-1} \mu' \,, \label{NS-ib-4}
\end{eqnarray}
which give rise to the following first integrals of motion
\begin{eqnarray}
& & b^2 = \frac{2 m B' }{B^2} \left( \frac{a'}{a} + \frac{2 M'}{M} \right)\,, \label{fint-ib-0} \\
& & I_1 =  f_1 m (m-1) \frac{a}{b} M^2 B^{m-1} \left[ \frac{4 b}{M} - 2 \left( \frac{a'}{a} +  \frac{2 M'}{M} \right) + (2 m -1) \frac{B'}{B}  \right] \,, \label{fint-ib-1}   \\ & & I_2 = - \frac{2 f_1 m (m-1)}{a b M^2}  B^{m-2} B' \, ,  \label{fint-ib-2} \\ & & I_3 = 2 f_1 m (m-1) \frac{a}{b} M^2 B^{-m}  \left( \frac{2 b}{M} - \frac{a'}{a} - \frac{2 M'}{M} \right)\,, \label{fint-ib-3}  \\ & & I_4 = - \frac{2 f_1 m (m-1) \mu(r)}{a b M^2}  B^{m-2}  B' \, ,  \label{fint-ib-4}
\end{eqnarray}
where $\mu(r)$ is an integration function. Firstly, it follows from \eqref{fint-ib-2} and \eqref{fint-ib-4} that $\mu (r) = I_4 / I_2$ is a constant ($I_2 \neq 0$). Then, we find from the above first integrals that
\begin{eqnarray}
& & k_1^2 \left[ I_1 + (2m -1) \frac{I_2}{2} a^2 M^4 \right] B - I_3 a^2 M^2 = 0\,, \label{fint-ib0-0} \\ & & a  = k_1 \frac{B^m}{M} \, ,  \label{fint-ib1-1} \\ & & \frac{B'}{B^2} = -\frac{k_1 I_2 b M }{ 2 f_1 m (m-1)} \, ,  \label{fint-ib2-2} \\ & & \frac{a'}{a} + \frac{2 M'}{M} =  -\frac{f_1 (m-1) b }{ k_1 I_2 M } \, ,  \label{fint-ib3-3}
\end{eqnarray}
where $k_1$ is defined by $k_1 \equiv I_3/(4 f_1 m(m-1)) - f_1 (m-1)/(2 I_2)$.
Now, by considering the relation \eqref{fint-ib3-3} in the definition of the boundary term $B$ given by \eqref{B2}, we find
\begin{eqnarray}
& & B = \frac{2}{M^2}\left[ \frac{f_1 (m-1)}{k_1 I_2} +2 \right] \left[ \frac{M'}{b} + \frac{f_1 (m-1)}{k_1 I_2} \right] \, , \label{B3}
\end{eqnarray}
for $I_3 \neq 0$. Then, using the relations \eqref{fint-ib1-1}-\eqref{fint-ib3-3}, we also find that
\begin{eqnarray}
& & B = \frac{2 f_1 (m-1)}{k_1 I_2 M^2} \left[ \frac{3 M'}{b} + \frac{ f_1 (m-1)}{k_1 I_2} \right]. \label{B4}
\end{eqnarray}
By comparing \eqref{B3} and \eqref{B4}, it is explicitly seen that the boundary term $B$ obtained from the definition does not coincide with the one getting from the Eqs. \eqref{fint-ib1-1}-\eqref{fint-ib3-3}. In order to show this contradiction between the statements of $B$, let us give an example in isotropic coordinates, i.e. we take $M(r) = r b(r)$ which yields $b(r) = (r/r_0)^3$ from \eqref{feq-ia-2-2}, where $r_0$ is an integration constant. Then, Eq. \eqref{fint-ib3-3} becomes
\begin{equation}
a(r) = \left( \frac{r}{r_1} \right)^{-8 - \frac{f_1 (m-1)}{k_1 I_2}}\,,  \label{a-sol-ib}
\end{equation}
where $r_1$ is an integration constant. Furthermore, from Eq.~\eqref{fint-ib2-2} we find that 
\begin{equation}
B(r) = \frac{16 r_0^6 f_1 m (m-1)}{k_1 I_2 r^8} \, .  \label{B5}
\end{equation}
Putting those of the results into the field equations \eqref{feq-ia-1}-\eqref{feq-ia-3}, all of them are satisfied under the condition $f_1= 2 k_1 I_2 ( 4m -3)/(m-1)$, which gives that  
\begin{equation}
a(r) = \left( \frac{r}{r_1} \right)^{-2 - 8 m}, \qquad B(r) = \frac{32 r_0^6 m (4m-3)}{r^8} \, .  \label{a-sol-B6}
\end{equation}
But, the definition of the boundary term $B$ given by \eqref{B2} yields
\begin{equation}
B(r) = \frac{16 r_0^6 (2m -1)(4m-1)}{r^8} \, .  \label{B6}
\end{equation}
Here we have explicitly exposed a contradiction between the statements of $B$. Therefore, we conclude that there is only one consistent interpretation for this case that it should be $m=1$, and there is no solution for $m \neq 1$. Let us clarify this here. We cannot argue that there are no exact solutions for this model but instead, we can say that using the Noether's symmetry approach, one cannot find solutions.

If $m=1/2$, then the Noether symmetry ${\bf X}_3$ is the only one which is different, having the form
\begin{eqnarray}
& & {\bf X}_3 = \frac{1}{2} \left( 2 \ln B -3 \right) a \partial_a - \frac{1}{2} \left( \ln B -1 \right) b \partial_b -  \frac{M}{2} \left( \ln B -1 \right) \partial_M + B \ln B \partial_B  \qquad {\rm with} \quad K= 2 f_1 \frac{a M }{\sqrt{|B|}}\,,   \label{NS-ib-3-2}
\end{eqnarray}
and the first integral related with ${\bf X}_3$ is
\begin{eqnarray}
& & I_3 = f_1 \frac{a M^2 }{2 b \sqrt{|B|}}  \left(\ln B \left[  \frac{a'}{a} + \frac{2 M'}{M} -\frac{2 b}{M}  \right] - \frac{B'}{2 B} \right) \,.  \label{fint-ib-3-2}
\end{eqnarray}
The remaining Noether symmetries are the same as ${\bf X}_0, {\bf X}_1, {\bf X}_2$ and ${\bf X}_4$ together with the first integrals \eqref{fint-ib-0}, \eqref{fint-ib-1}, \eqref{fint-ib-2} and \eqref{fint-ib-4}, but with $m= 1/2$. After some algebra, we have also determined a contradiction between the statements of $B$ for $m=1/2$, which means that there is no any consistent solution using the Noether's integrals in this case.

    \paragraph{Subcase: $f(T,B)=f_0 T^n + f_1 B^n$ ($n= m$)}
For this case, there are again {\it two} Noether symmetries, ${\bf X}_0$ and
\begin{eqnarray}
& & {\bf X}_1 = (2n-3) a \partial_a +  b \partial_b + M \partial_M - 2 T \partial_T - 2 B \partial_B \,. \label{X1-4}
\end{eqnarray}
The first integrals for ${\bf X}_0$ and ${\bf X}_1$ give the following relations
\begin{eqnarray}
& & b^2 = \frac{ M^2 \left[ \frac{2 a' M'}{a M} + \frac{M'^2}{M^2} - \frac{(n-1) f_1}{f_0} \left( \frac{B}{T} \right)^{n-1} \frac{B'}{B} \left( \frac{a'}{a} + \frac{2 M'}{M} \right) \right] }{ 1 + \frac{(1-n)}{2n} M^2 T \left[ 1 + \frac{f_1}{f_0} \left( \frac{B}{T} \right)^n \right] }\,,   \label{NC1-4} \\& & \frac{I_1 b \, T^{1-n}}{4 f_0 n a M^2} =   \frac{a'}{a} +  2 (n-1) \left( \frac{M'}{M} - \frac{b}{M} \right)  \nonumber \\& & \qquad \qquad  \qquad + (n-1) \frac{f_1}{2 f_0} \left( \frac{B}{T} \right)^{n-1} \left[ 2 \left( \frac{a'}{a} + \frac{2 M'}{M} \right) + (1-2n) \frac{B'}{B}  - \frac{4 b}{M} \right]\, .   \label{NC2-4}
\end{eqnarray}

Considering the definition of $T$ in Eq.~\eqref{NC1-4}, we find
\begin{eqnarray}
& & (1 - 2n) \left( \frac{2 a' M'}{a M} + \frac{M'^2}{M^2} \right) +  2 (n-1)\frac{b}{M} \left( \frac{a'}{a} + \frac{M'}{M} \right) + \frac{b^2}{M^2}  \nonumber \\ & & \qquad \qquad \qquad \qquad \qquad \qquad   + \frac{(1-n) f_1 T}{f_0} \left( \frac{B}{T} \right)^{n} \left[ \frac{b^2}{2} - \frac{n B'}{B^2} \left( \frac{a'}{a} + \frac{2 M'}{M} \right)  \right] = 0\,. \label{ceq-i-2}
\end{eqnarray}
Now, we take $M(r)=r$, which simplifies the above first integrals to
\begin{eqnarray}
& & 2 \left[ 1 - 2n + (n-1) b \right] \frac{a'}{a} + \frac{ \left[ 1 -2n + 2(n-1) b + b^2 \right]}{r} +  \frac{(1-n) f_1 r T}{2 f_0} \left( \frac{B}{T} \right)^{n} \left[ b^2 - \frac{2 n B'}{B^2} \left( \frac{a'}{a} + \frac{2}{r} \right)  \right] = 0\,, \label{ceq-i-2-Mr1} \\ & &  \frac{a'}{a} + \frac{ 2(n-1)(1- b)}{r} +  \frac{(n-1) f_1 }{2 f_0} \left( \frac{B}{T} \right)^{n-1} \left[ \frac{2 a'}{a}  +\frac{4(1-b)}{r} +(1 - 2 n) \frac{B'}{B} \right] - \frac{I_1 b T^{1-n}}{4 f_0 n r^2 a} = 0\,, \label{ceq-i-2-Mr2}
\end{eqnarray}
Here, the field equations \eqref{var-a-2}, \eqref{var-b-2} and \eqref{var-M-2} have the following form 
\begin{eqnarray}
& & \frac{2}{r^2 b^2} \left( 2 r \frac{b'}{b} + b^2 -1 \right) = \rho_{_{TB}}, \label{feq-i-2-1} \\ & & \frac{2}{r^2 b^2} \left( 2 r \frac{a'}{a} - b^2 +1 \right) = p_{_{TB}}\,, \label{feq-i-2-2}
\end{eqnarray}
and
\begin{eqnarray}
& & \frac{a''}{a} - \frac{a' b'}{a b} + \frac{1}{r} \left( \frac{a'}{a} + \frac{b'}{b} \right)  + \frac{(b^2 -1)}{r^2} = (1-n) \left\{  \left( \frac{a'}{a} + \frac{b-1}{r} \right) \frac{T'}{T} + \frac{f_1 b}{f_0 r} \left( \frac{B}{T} \right)^{n-1} \frac{B'}{B}  \right\} \, , \label{feq-i-2-3}
\end{eqnarray}
where $\rho_{_{TB}}$ and $p_{_{TB}}$ become
\begin{eqnarray}
& & \rho_{_{TB}} = \frac{(n-1)}{n} T  + \frac{4 (n-1) (1-b)}{r\, b^2} \frac{T'}{T}  +  \frac{2(n-1) f_1}{f_0 \, b^2} \left( \frac{B}{T} \right)^{n-1} \left[  \frac{b^2 B}{2 n} + \left( \frac{b'}{b} - \frac{2 b}{r} \right) \frac{B'}{B} - (n-2) \frac{B'^2}{B^2} - \frac{B''}{B} \right]  \, , \qquad \label{rho-i-2} \\ & & p_{_{TB}} = \frac{(1-n)}{n} T +  \frac{(1-n) f_1 }{f_0 \, b^2} \left( \frac{B}{T} \right)^{n-1} \left[ \frac{b^2 B}{n} + \frac{2 B'}{B} \left( \frac{a'}{a} + \frac{2}{r} \right) \right] \, .
\end{eqnarray}
 For this case, we can get a solution of the field equations when $I_1 = 0$, which is
\begin{eqnarray}
& & a(r) = \left( \frac{r}{r_0} \right)^q, \qquad b(r) = k\,,  \label{sol-case-1d}
\end{eqnarray}
where $r_0$ is an integration constant and $k$ is a constant. This solution has the following extra relationship between the parameters,
\begin{eqnarray}
& & n =  \frac{ (k+2)(k-1 +q) \pm \tilde{C} }{ 4( k -1 + q) } \, , \qquad \,\, \\& & f_1 = \frac{ f_0 (k-1) [ 4 n^2 - 2 n (q+3) + 2 q -k +1 ]}{(n-1) [ 4 n^2 - 2 n ( 2k + 1) + (q+1) ( 2k-q-2)]} \left[ \frac{ (q+1) (2 k -q-2) }{(k-1)( 2 q -k+1)} \right]^{1-n} \, ,\label{f1is}\\
&& \tilde{C}=\sqrt{ 4 q^4 + 4 (2-k) q^3 -3( k^2 -8k +4) q^2 - 2(k-1)(k+2) (k-4) q -(k-1)^2 (7 k^2 -16)}\,,
\end{eqnarray}
where $k, q$ are constant parameters. Then, the quantities $T(r), B(r), \rho_{_{TB}} $ and $p_{_{TB}}$ for the above solution are
\begin{eqnarray}
& & T(r) = \frac{T_0}{r^2}\,, \qquad B(r) = \frac{B_0}{r^2}\,, \qquad  p_{_{TB}} = \frac{ p_0}{r^2} \, , \qquad \rho_{_{TB}} = \frac{\rho_0}{r^2} \, ,
\end{eqnarray}
where $T_0, B_0, p_0$ and $\rho_0$ have the following form
\begin{eqnarray}
& & T_0 = \frac{2 (k-1)(k-1 -2q)}{k^2}\,, \qquad B_0 = \frac{2 (q+1)(q+2 -2k)}{k^2}\,,   \\ & &   p_0 = \frac{2 (k-1)}{ n k^2} \left[ (1-n)( k - 2 q -1)  - \frac{ [ 4 n^2 - 2 n (q+3) + 2 q -k + 1 ] [ (q+1) (q- 2 k +2) - 2 n ( q +2) ] }{[ 4 n^2 - 2 n (2k +1) + (q +1) (2 k - q -2) ]} \right] \, , \\ & & \rho_0 = \frac{ 2 ( k^2 - 1)}{ k^2}  \, .
\end{eqnarray}
For this case, the equation of state $w_{_{TB}} = p_0 / \rho_0$ becomes 
\begin{eqnarray}
& &  w_{_{TB}} = \frac{1}{ n (k+1)} \left[ (1-n)( k - 2 q -1)  - \frac{ [ 4 n^2 - 2 n (q+3) + 2 q -k + 1 ] [ (q+1) (q- 2 k +2) - 2 n ( q +2) ] }{[ 4 n^2 - 2 n (2k +1) + (q +1) (2 k - q -2) ]} \right] \, . \quad \label{w-case-1d}
\end{eqnarray}

\subsubsection{Case 2: $ f(T,B)= T\, h_1 (B) + h_2 (B)$ ($f_{TT} = 0$)}
When one considers the condition $f_{TT} = 0$, there are several subcases having different Noether's symmetries. In the following, we will study them separately.  
    \paragraph{Subcase: $ f(T,B)= f_0 T B^m$ ($h_1 (B) = f_0 B^m$ and $h_2 (B) = f_1 B$)}
As it happened before, the case $h_2 (B) = f_1 B$ is a special case where the boundary term does not affect the field equations since only appears linearly. Then, this subcase is identical as choosing $f_1=0$. This theory admits the Noether symmetries ${\bf X}_0$ and
\begin{eqnarray}
& & {\bf X}_1 = (2m -1) a \partial_a + b \partial_b + M \partial_M - 2 T \partial_T - 2 B \partial_B \quad {\rm with} \quad K= -8 f_1 m a M \, ,  \label{ii-X1-2}
\end{eqnarray}
and the corresponding first integrals are
\begin{eqnarray}
& & b^2 = \frac{1}{1 - \frac{m}{2} M^2 T} \left\{  M M' \left( \frac{2 a'}{a} + \frac{M'}{M} \right) - \frac{m M^2 T}{B} \left(  \frac{a'}{a} + \frac{2 M'}{M} \right) \left[  \frac{T'}{T} + (m-1) \frac{B'}{B} \right] \right\} \,,  \label{fint-ii-1} \\ & &  I_1  = - 4 f_0 a M T B^{m-1} \left[ 2 m^2 + (2m -1) \frac{B}{T}  + \frac{m^2 M}{b} \left( \frac{a'}{a} + \frac{2 M'}{M} \right) + m ( 2m +1) \frac{M}{2 b} \left( \frac{T'}{T} + (m-1) \frac{B'}{B} \right) \right] \,, \qquad  \label{fint-ii-2}
\end{eqnarray}
where $T \neq 2/ (m M^2)$. 
For $M(r)= r$ and after considering the definition of $T$, the above first integrals become
\begin{eqnarray}
& &  2 \left[ 1+ m (1 -b) \right] \frac{a'}{a} + \frac{(m+1)(1 - b)}{r} - m \frac{r \, T}{B} \left(  \frac{a'}{a} + \frac{2}{r} \right) \left[  \frac{T'}{T} + (m-1) \frac{B'}{B} \right] = 0 \,,  \label{fint-ii-1-2} \\ & &  \frac{a'}{a} +  \frac{2 (1+ b)}{r} + (2m -1) \frac{b B}{m^2 r T}  +  \frac{( 2m +1)}{2 m} \left[ \frac{T'}{T} + (m-1) \frac{B'}{B} \right]  + \frac{I_1 b B^{m-1}}{4 f_0 m^2 r^2 a T} = 0 \, .   \label{fint-ii-2-2}
\end{eqnarray}
In this case, it is difficult to find solutions using the Noether's symmetry approach. Moreover, even for the case when $a(r) = \left( r/r_0 \right)^q$ and $b(r) = k = const.$, there are no solutions.

    \paragraph{Subcase: $f(T,B)= f_0 T B^{1/2}$ ($h_1(B) = f_0 B^{1/2}$ and $h_2 (B) = f_1 B=0$)}
  In this case, the Noether symmetries are ${\bf X}_0$ and
\begin{eqnarray}
& & {\bf X}_1 = b \partial_b + M \partial_M - 2 T \partial_T - 2 B \partial_B \quad {\rm with} \quad K= -4 f_1 a M \label{ii-X1-1}
\end{eqnarray}
which can be used with the first integrals to get the following two equations,
\begin{eqnarray}
& & b^2 = \frac{M^2}{1 - \frac{M^2}{4} T} \left[  \frac{M'}{M} \left( \frac{2 a'}{a} + \frac{M'}{M} \right) - \frac{T}{2 B} \left(  \frac{a'}{a} + \frac{2 M'}{M} \right) \left(  \frac{T'}{T} - \frac{B'}{2 B} \right) \right] \,,  \label{fint-ii-3} \\ & & I_1 = - \frac{2 f_0 a M^2 T}{b \sqrt{B}} \left[ \frac{b}{M} + \frac{1}{2} \left( \frac{a'}{a} + \frac{2 M'}{M} \right) +   \frac{T'}{T} - \frac{B'}{2 B} \right]  \, .\label{fint-ii-4}
\end{eqnarray}
By using the definition of $T$ and inserting $M(r) = r$ in \eqref{fint-ii-3} and \eqref{fint-ii-4}, we obtain that 
\begin{eqnarray}
& & 2 (3 - b) \frac{a'}{a} + \frac{( 3- 2 b - b^2)}{r} - \frac{r \, T}{B} \left(  \frac{a'}{a} + \frac{2}{r} \right) \left(  \frac{T'}{T} - \frac{B'}{2 B} \right) = 0\, ,  \label{fint-ii-5} \\ & &   \frac{a'}{a} + \frac{2 (1 + b)}{r} + \frac{2 T'}{T} - \frac{B'}{B} + \frac{I_1 b \sqrt{B}}{f_0 a r^2 T} = 0  \, .\label{fint-ii-6}
\end{eqnarray}
If $I_1 = 0$, Eqs.~\eqref{fint-ii-5} and \eqref{fint-ii-6} take the following forms
\begin{eqnarray}
& & 2 (3 - b) \frac{a'}{a} + \frac{( 3- 2 b - b^2)}{r} + \frac{r \, T}{2 B} \left(  \frac{a'}{a} + \frac{2}{r} \right) \left[  \frac{a'}{a} + \frac{2( 1 + b)}{r} \right] = 0 \, ,  \label{fint-ii-5-2}\\ & &   \frac{a'}{a} + \frac{2 (1 + b)}{r} + \frac{2 T'}{T} - \frac{B'}{B} = 0  \, . \label{fint-ii-6-2}
\end{eqnarray}
Similarly as the previous case, it is also hard to get exact solutions for this case and there are no solutions for the case $a(r) = \left( r/r_0 \right)^q$ and $b(r) = k = const$.

\paragraph{Subcase: $f(T,B)= f_0 T + f_1 B + f_2 B \ln B $ ($h_1 (B) = f_0 + f_1 B / T$ and $h_2 (B) = f_2 B \ln B$)}
 For this case, the Noether symmetries are ${\bf X}_0$ and
\begin{eqnarray}
& & {\bf X}_1 = - a \partial_a + b \partial_b + M \partial_M  - 2 B \partial_B \quad {\rm with} \quad K= 8 f_2 a M \, ,  \label{ii-X1-3}
\end{eqnarray}
which have the first integral
\begin{eqnarray}
& & I_1 = \frac{4 a M^2}{b} \left[ f_0 \frac{a'}{a} + f_2 \left( \frac{a'}{a} + \frac{2 M'}{M} - \frac{2 b}{M} - \frac{B'}{2 B} \right) \right]\,, \label{fint-ii-3-2}\\
& & b^2 = \frac{ M^2}{ f_0 - \frac{f_2}{2} M^2 B } \left[  f_0 \frac{M'}{M} \left(  \frac{2 a'}{a} + \frac{M'}{M} \right) - f_2 \left( \frac{a'}{a} + \frac{2 M'}{M} \right) \frac{B'}{B} \right] \label{fint-ii-3-1} \,.
\end{eqnarray}
Now, assuming $M(r) = r$ in the above first integrals, we have
\begin{eqnarray}
& &  \frac{2 a'}{a} + \frac{(1- b^2)}{r} + \frac{f_2 r}{f_0} \left[ \frac{b^2}{2} B -  \left( \frac{a'}{a} + \frac{2}{r} \right) \frac{B'}{B}  \right] = 0 \,,  \label{fint-ii-3-1-2} \\ & & 2 (f_0 + f_2) \frac{a'}{a} + \frac{ 4 f_2 (1 - b) ]}{r}  - f_2 \frac{B'}{B} - \frac{I_1 b}{2 r^2 a} =0\,. \label{fint-ii-3-2-2}
\end{eqnarray}
Here, the field equations \eqref{var-a-2} and \eqref{var-b-2} have the same form as \eqref{feq-i-2-1} and \eqref{feq-i-2-2}, respectively,  in which $\rho_{_{TB}}$ and $p_{_{TB}}$ are given by
\begin{eqnarray}
& & \rho_{_{TB}} =  \frac{f_2}{f_0} \left\{ B + \frac{2}{b^2} \left[ \left( \frac{b'}{b} - \frac{2 b}{r} \right) \frac{B'}{B} - \frac{B''}{B} + \frac{B'^2}{B^2} \right] \right\} \, ,  \label{rho-ii-1} \\ & & p_{_{TB}} = - \frac{f_2}{f_0} \left[ B + \frac{2}{b^2} \frac{B'}{B} \left( \frac{a'}{a} + \frac{2}{r} \right) \right]  \, .
\end{eqnarray}
The third field equation (which is a combination of the other two) given in \eqref{var-M-2} has the following form
\begin{eqnarray}
& & \frac{a''}{a} - \frac{a' b'}{a b} + \frac{1}{r} \left( \frac{a'}{a} + \frac{b'}{b}  \right) + \frac{(b^2-1)}{r^2} +  \frac{f_2 b}{f_0 r} \frac{B'}{B}  = 0 \, . \label{feq-ii-3}
\end{eqnarray}
For $I_1 = 0$, we found the following solution
\begin{eqnarray}
& & a(r) = \left( \frac{r}{r_0} \right)^q \, , \qquad b(r) = k, \label{sol-a-case-2c}
\end{eqnarray}
where $r_0$ is an integration constant, and the parameters must have the following form:
\begin{eqnarray}
& & k = \frac{f_2 \left( f_0^2 + 8 f_0 f_2 + f_2^2 \right) \pm (f_0+f_2) K_1 }{f_0 \left(f_0^2+2 f_0 f_2+5 f_2^2\right)}\, , \label{k-case-2c}\\
& & q = \frac{f_2 \left[ (f_0 + f_2)(2 f_2 - 3 f_0) \pm 2 K_1 \right]}{f_0 \left(f_0^2+2 f_0 f_2+5 f_2^2\right)} \, ,
\end{eqnarray}
where $ K_1^2 =f_0^4+2 f_0^3 f_2-3 f_0^2 f_2^2+14 f_0 f_2^3+f_2^4$. For the solution \eqref{sol-a-case-2c}, $T(r), B(r), p_{_{TB}}$ and $\rho_{_{TB}}$ become \begin{eqnarray}
& & T(r) = \frac{2 (k-1) ( k-1 -2q)}{ k^2 r^2} \,, \quad B(r) = \frac{ 2 (q+1)(q-2k +2)}{k^2 r^2} \, , \label{T-B-case-2c} \\ & & p_{_{TB}} = \frac{2 f_2\left[ 2 k (q+1)- (q-1)(q+2) \right]}{k^2 f_0 r^2}\,, \quad \rho_{_{TB}} = \frac{2 f_2[ q (q+3)-2 k (q-1)]}{k^2 f_0 r^2}\, , \label{p-rho-case-2c}
\end{eqnarray}
in which the latter $p_{_{TB}}$ and $\rho_{_{TB}}$ require that the equation of state has the form
\begin{eqnarray}
& & w_{_{TB}} = \frac{  2 k (q+1)- (q-1)(q+2) }{  q (q+3)-2 k (q-1) }\, .  \label{w-case-2c}
\end{eqnarray}


\subsubsection{Case 3: $f(T,B)= g_1 (T) + B g_2 (T)$ ($f_{BB} = 0$)}
By imposing $f_{BB} = 0$, the function $f(T,B)$ found from the Noether's symmetry equations (see Eqs.~\eqref{noether-eqs}) takes the form 
\begin{equation}
    f(T,B)= g_1 (T) + B g_2 (T)\,,
\end{equation}
where we can have two non-trivial functions that will be study separately. 

    \paragraph{Subcase: $f(T,B)=\tilde{f_0} T+ f_2 B \ln T$ ($g_1(T) = \tilde{f_0} T$ and $g_2 (T) = f_2 \ln T$)}
In this subcase, it is found that the Noether symmetries are ${\bf X}_0$ and
\begin{eqnarray}
& & {\bf X}_1 = - a \partial_a + b \partial_b + M \partial_M - 2 T \partial_T - 2 B \partial_B \quad {\rm with} \quad K= 8 f_2 a M \,. \label{iii-X1}
\end{eqnarray}
Hence the first integrals due to ${\bf X}_0$ and ${\bf X}_1$ are, respectively,
\begin{eqnarray}
& & b^2 = \frac{ M^2}{ \tilde{f_0} + f_2 B \left( \frac{1}{T} - \frac{M^2}{2} \right) } \left[  \left( \tilde{f_0} + f_2 \frac{B}{T} \right) \frac{M'}{M} \left(  \frac{2 a'}{a} + \frac{M'}{M} \right) - f_2 \left( \frac{a'}{a} + \frac{2 M'}{M} \right) \frac{T'}{T} \right],   \label{fint-iii-0} \\ & & I_1 = \frac{4 a M^2}{b} \left[ \left( \tilde{f_0} + f_2 \frac{B}{T} \right) \frac{a'}{a}  + f_2 \left( \frac{a'}{a} + \frac{2 M'}{M} - \frac{2 b}{M} - \frac{T'}{2 T} \right) \right].  \label{fint-iii-1}
\end{eqnarray}
Considering the $M(r) = r$ case, the above first integrals reduce to 
\begin{eqnarray}
& & \left( \tilde{f_0} + f_2 \frac{B}{T} \right) \left[ \frac{2 a'}{a} + \frac{(1-b^2)}{r}  \right] + f_2 r \left[ \frac{b^2}{2} B - \left( \frac{a'}{a} + \frac{2}{r} \right) \frac{T'}{T} \right] = 0\,, \label{iii-fint-2-1}  \\& & \left( \tilde{f_0}+ f_2 \frac{B}{T} \right) \frac{a'}{a } + f_2 \left[  \frac{a'}{a} + \frac{2 ( 1- b)}{r} - \frac{T'}{2 T} \right] - \frac{I_1 b}{4 r^2 a} = 0\,. \label{iii-fint-2-2}
\end{eqnarray}
When $I_1 = 0$, we obtained the following solution
\begin{eqnarray}
& & a(r) = \left( \frac{r}{r_0} \right)^q \, , \qquad b(r) = k\,, \label{sol-a-case-3a}
\end{eqnarray}
where $r_0$ is an integration constant, and there the parameters must be
\begin{eqnarray}
& & \tilde{f}_0 = \frac{ 2 f_2 \left[ (2 q^2 + 5 q + 4) k^2 - k (q+8)(q+1)^2 + (2 q +1) (q+2)^2  \right]}{(k-1) (k- 2q -1)( k^2 - 2q -1) } \, ,\label{f0tildeis}\\
& & k = \frac{K_2^2 + K_2 (q+3)-11 q^2+18 q+21}{6 K_2} \, ,\\
& & K_2^3 = (q+3) (37 q^2 + 24 q -27) + \sqrt{75 q^6+96 q^5+684 q^4+702 q^3-714 q^2-864 q-75} \,.
\end{eqnarray}
For the solution \eqref{sol-a-case-3a}, $T(r), \ B(r)$, $p_{_{TB}} $ and $\rho_{_{TB}}$ become 
\begin{eqnarray}
& & p_{_{TB}} = \frac{ 2 (2q -k^2 + 1) [ 2 k (q+1) - (q-1)(q+2)]}{ k^2 [ 2 k (q+1) - (q+1)(q+3)] r^2}\,, \quad \rho_{_{TB}} = \frac{ 2 ( k^2 -2 q - 1) [ 2 k (q-1) - q (q+3)]}{ k^2 [ 2 k (q+1) - (q+1)(q+3)] r^2}\,,      \label{p-rho-case-3a}
\end{eqnarray}
which gives the equation of state $w_{_{TB}} = p_{_{TB}} / \rho_{_{TB}}$ as:
\begin{eqnarray}
& & w_{_{TB}} = -\frac{ [ 2 k (q+1) - (q-1)(q+2)]}{  2 k (q-1) - q (q+3) } \, .   \label{w-case-3a}
\end{eqnarray}

    \paragraph{Subcase: $f(T,B)=f_1 B T^n$ ($g_1(T) = 0$ and $g_2 (T) = f_1 T^n$)}
 Here there exists two Noether symmetries ${\bf X}_0$ and
\begin{eqnarray}
& & {\bf X}_1 = (2n -1) a \partial_a + b \partial_b - M \partial_M - 2 T \partial_T - 2 B \partial_B\,, \label{iv-X1-1}
\end{eqnarray}
which have the first integrals
\begin{eqnarray}
& & b^2 = \frac{M^2}{1 - \frac{M^2}{2} T} \left[  \frac{M'}{M} \left(  \frac{2 a'}{a} + \frac{M'}{M} \right) - \frac{T'}{B} \left( \frac{a'}{a} + \frac{2 M'}{M} \right) \right]\,,   \label{iv-fint-1} \\ & & I_1 = 4 f_1 n T^n \frac{a M^2}{b} \left\{  \frac{a'}{a} + \frac{2 M'}{M} - \frac{2 b}{M} + \frac{B}{T} \left[ \frac{a'}{a} + 2n \left( \frac{M'}{M} - \frac{b}{M} \right) \right] - (2n +1) \frac{T'}{2 T}  \right\}\,. \label{iv-fint-2}
\end{eqnarray}
Using the definition of $T$ in \eqref{iv-fint-1} yields
\begin{eqnarray}
& & \frac{2 a' M'}{a M } + \frac{M'^2}{M^2} - \frac{b}{M} \left( \frac{a'}{a} + \frac{M'}{M} \right) - \frac{T'}{2 B} \left( \frac{a'}{a} + \frac{2 M'}{M} \right) = 0\,. \label{iv-fint-1-2}
\end{eqnarray}
Again we restrict the analysis to the $M(r) = r$ case. Then, the above Eqs. \eqref{iv-fint-2} and \eqref{iv-fint-1-2} take the forms
\begin{eqnarray}
& & \frac{a'}{a} + \frac{2 (1-b)}{r} + \frac{B}{T} \left[ \frac{a'}{a} + \frac{2n (1-b)}{r} \right] - (2n +1) \frac{T'}{2 T} - \frac{I_1 b T^{-n}}{ 4 f_1 n r^2 a} = 0, \label{iv-fint-2-2}  \\& & (2 - b) \frac{a'}{a } + \frac{(1-b)}{r} - \frac{r \, T'}{2 B} \left( \frac{a'}{a} + \frac{2}{r} \right) = 0. \label{iv-fint-1-3}
\end{eqnarray}

When $I_1 = 0$, the solution for this case has the form 
\begin{eqnarray}
& & a(r) = \left( \frac{r}{r_0} \right)^q \, , \qquad b(r) = k\,, \label{sol-a-case-3b}
\end{eqnarray}
where $r_0$ is an integration constant, and $k, n$ have the the following form 
\begin{eqnarray}
& & k =  \frac{ ( q+8)( q+1)^2 \mp q C }{ 2 ( 2 q^2 + 5 q + 4) }\,, \label{k-case-3b}  \\ & & n = \frac{ (q+1)( 3 q^3 + 9 q^2 + 20 q + 16) \pm q^2 C }{4 (q+2) \left(2 q^2 + 5 q + 4 \right)} \, , \label{n-case-3b}
\end{eqnarray}
where $C^2 = q^4 + 4 q^3 + 22 q^2 + 48 q + 33 $. For this solution, $p_{_{TB}}$ and $\rho_{_{TB}}$ have the form
\begin{eqnarray}
& & p_{_{TB}} = \frac{ p_0 }{ r^2 }\,,  \qquad \rho_{_{TB}} = \frac{\rho_0 }{ r^2} \, , \label{rho-p-case-3b}
\end{eqnarray}
where $p_0$ and $\rho_0$ are given by 
\begin{eqnarray}
& & p_0 = \frac{ 2 q ( q^2 + 6 q + 7 \mp C)( 7 q^2 + 14 q + 9 \pm C) \left[ q^4 - 4 q^3 -22 q^2 -31 q -16 \pm q (q+1) C \right] }{ (q+1)( q^2 -q -3 \pm C)[ (q+8)(q+1)^2 \mp q C]^2 }\,,  \\ & & \rho_0 = \frac{ 2 q ( q^2 + 6q + 7 \mp C) \left[ 9 q^4 - 12 q^3 - 101 q^2 - 130 q - 48 \pm ( 15 q^2 + 30 q + 16) C  \right] }{ (q^2 - q -3 \pm C) [ (q+8)(q+1)^2 \mp q C]^2} \, .
\end{eqnarray}
Here, the equation of state $w_{_{TB}} = p_0 / \rho_0$ becomes
\begin{eqnarray}
& & w_{_{TB}} = \frac{ ( 7 q^2 + 14 q + 9 \pm C) \left[ q^4 - 4 q^3 -22 q^2 -31 q -16 \pm q (q+1) C \right] }{ (q+1) \left[ 9 q^4 - 12 q^3 - 101 q^2 - 130 q - 48 \pm ( 15 q^2 + 30 q + 16) C  \right]  }  \, ,
\end{eqnarray}
where $q \neq -1$.

\subsubsection{Case 4: $f_{TB} \neq 0$}
For this case, we find that the Noether's symmetry equations gives us that the form of the function $f(T,B) = f_1 T^n B^m$, and we can obtain different symmetries depending on the parameters $n$ and $m$.
    \paragraph{Subcase: $f(T,B) = f_1 T^n B^{1-n}$  ($n$ arbitrary, $m = 1-n$)}
In this case, there are again two Noether symmetries ${\bf X}_0$ and
\begin{eqnarray}
& & {\bf X}_1 = - a \partial_a + b \partial_b + M \partial_M + \eta^4(r,a,b,M,T,B) \partial_T + \frac{B}{T} \eta^4 \partial_B \, , \label{iv-X1-2}
\end{eqnarray}
where $\eta^4$ is an arbitrary function of $r,a,b,M,T$ and $B$. Then, the Noether constants for these vector fields are
\begin{eqnarray}
& & b^2 = M^2 \left[ \frac{M'}{M} \left( \frac{2 a'}{a} + \frac{M'}{M} \right) + (n-1) \frac{T}{B} \left( \frac{a'}{a} + \frac{2 M'}{M} \right) \left( \frac{T'}{T} - \frac{B'}{B} \right) \right]\,, \label{fint-ivb-1} \\ & & I_1 = 2 f_1 n \frac{ a M^2}{b} \left( \frac{T}{B} \right)^{n-1}  \left[  \frac{2 a'}{a} + (n-1) \frac{T}{B} \left( \frac{T'}{T} - \frac{B'}{B} \right) \right]\,. \label{fint-ivb-2}
\end{eqnarray}
Because of the arbitrariness of $\eta^4$ which means that there are infinitely many Noether symmetries, one can choose $\eta^4 =0$, without loss of generality. Now, by taking $M(r) = r$, Eqs. \eqref{fint-ivb-1} and \eqref{fint-ivb-2} become
\begin{eqnarray}
& &  \frac{2 a'}{a} + \frac{(1-b^2)}{r} + (n-1) \frac{r T}{B} \left( \frac{a'}{a} + \frac{2}{r}  \right) \left( \frac{T'}{T} - \frac{B'}{B} \right) = 0\,, \label{fint-ivb-1-1} \\ & & \frac{2 a'}{a} + (n-1) \frac{T}{B} \left( \frac{T'}{T} - \frac{B'}{B} \right) - \frac{I_1 b }{2 f_1 n r^2 a} \left( \frac{T}{B} \right)^{1-n} = 0\,. \label{fint-ivb-2-2}
\end{eqnarray}
The metric coefficient $b(r)$ can be directly solved If $I_1 = 0$. In this case, by replacing Eq.~\eqref{fint-ivb-2-2} into \eqref{fint-ivb-1-1} we find  
\begin{eqnarray}
b(r)&=&\frac{1}{a}\sqrt{-2 r^2 a'^2-2 r a(r) a'+a^2} \label{fint-ivb-2-3}
\end{eqnarray}
which can be used in Eq.~\eqref{fint-ivb-1} to get a third order differential equation for $a(r)$ that is not easy to find solutions. The easiest way to solve this in a non-trivial way is by assuming that $a(r)$ is a power-law. By doing this, one finds that the only non-trivial solution is when $b(r)=1/3$ and $a(r)=a_0 r^{-4/3}$ and $n<0$. However, this solution gives $B=0$, so that this solution is a trivial solution. 

    \paragraph{Subcase: $f(T,B) = f_1 T^n B^m$ ($n + m \neq 3/2$)}
In this case, we have two Noether symmetries ${\bf X}_0$ and
\begin{eqnarray}
& & {\bf X}_1 = (2n + 2m -3) a \partial_a + b \partial_b + M \partial_M - 2 T \partial_T - 2 B \partial_B \,  , \label{iv-X1-4}
\end{eqnarray}
which have the constants of motion
\begin{eqnarray}
& & b^2 = \frac{ M^2}{ n + (1-n-m) \frac{M^2}{2} T } \left[ \frac{n M'}{M} \left( \frac{2 a'}{a}  + \frac{M'}{M} \right) - m  \frac{T}{B} \left( \frac{a'}{a} + \frac{2 M'}{M} \right) \left( n \frac{ T'}{T} + (m-1)\frac{B'}{B} \right) \right] \, ,  \label{fint-iv-c-1}  \\ & & I_1 = 4 f_1 T^{n-1} B^m \frac{a M^2}{b} \Big\{ n \left[ \frac{a'}{a} + 2 ( n + m -1) \frac{M'}{M} \right] + 2 (n-m) (n + m - 1) \frac{b}{M}  \nonumber \\ & & \qquad \qquad \qquad \qquad \qquad  + \frac{m T}{2 B} \left[ 2 (n + m -1) \left( \frac{a'}{a} + \frac{2 M'}{M} \right) - (2 n + 2 m -1)  \left( n \frac{ T'}{T} + (m-1)\frac{B'}{B}  \right)  \right] \Big\} \,.  \label{fint-iv-c-2}
\end{eqnarray}
For $M(r)= r$, using the definition of $T$, the above equations reduce to 
\begin{eqnarray}
& & 2 \left[ n + \ell (2 - b) \right] \frac{a'}{a} +  \frac{ \left[ n + 2 \ell ( 1 - b) \right]}{r}  - m \frac{r T}{B} \left( \frac{a'}{a} + \frac{2}{r} \right) \left[ \frac{n T'}{T} + (m-1) \frac{B'}{B} \right] = 0\,,  \\ & & n \frac{a'}{a} + 2 \ell \frac{[ n + (n-m) b]}{r} + \frac{m T}{2 B} \Big{\{} 2 \ell \left( \frac{a'}{a} + \frac{2}{r} \right) - (2 \ell + 1) \left[ \frac{n T'}{T} + (m-1) \frac{B'}{B} \right] \Big{\}} - \frac{ I_1 b T^{1-n} }{4 f_1 r^2 a B^m} = 0\,,
\end{eqnarray}
where $\ell = n + m -1$.

\section{\uppercase{Conclusions}}\label{conclusions}
In the recent literature~\cite{Dialektopoulos:2018qoe,Bahamonde:2016jqq}, some examples where the Noether symmetry approach has been used as a geometric criterion to select theories of gravity in which the Noether symmetry generator constraints arbitrary functions in the action, and also allows to find out exact solutions for the field equations of a given gravity theory due to the conserved quantities, i.e. the first integrals of motion. It is either possible to study spherically symmetry in a gauge where $M(r)=r$ without setting $a(r)$ and $b(r)$, or  study with a generic $M(r)$ and setting the gauge $a(r)=1/b(r)$~\cite{Weinberg:1972kfs}. Even though both approaches should give rise to the same physics, the field equations are mathematically different. Since we were interested in finding exact solutions with the Noether's approach, we chose to work in full generality and in the end, see if it is convenient (for finding solutions) to work in one gauge or the other. In our study, we have found that the gauge $M=r$ was better than $a(r)=1/b(r)$ for this purpose since it is easier to find exact solutions \\

In this work, we have studied $f(T,B)$ gravity which is a modified Teleparallel theory of gravity where the manifold is flat but contains torsion. In this theory, $T$ is the torsion scalar and $B$ is the boundary term which is connected with the Ricci scalar as $\bar{R}=-T+B$. We have found the admitting Noether symmetries in the background of spherically symmetric space-time. In order to write out the Noether symmetry equations, we derived the point-like Lagrangian \eqref{L}, which gives rise to the dynamical field equations varying
with respect to the configuration space variables $a, b$ and $M$. After solving the Noether symmetry equations (in vacuum) and using the first integrals for the appropriate Noether symmetry, we have found some exact spherically symmetric solutions for several forms of the function $f(T,B)$ in different cases. The exact solutions found are displayed in Table~\ref{Table1}. Unfortunately, due to the difficulty of the field equations, we have only found solutions where the metric coefficient $b(r)$ is a constant. In the cases including power laws of the boundary term $B$, we also observe that there is no solution in which the metric coefficient $b(r)$ is a constant. For all the solutions that we found, if one tries to go to the GR limit, for example in the $f_0 T^n$ case, if one sets $n=1$ , one only gets a trivial metric, i.e., $a(r)=1$ or $0$ or $b(r)=1$ or $0$. Thus, all of these solutions are branching off the Schwarzschild solution and they become different from trivial when the modification of GR is considered.\\

It is important to mention that we have not assumed the form of the function $f(T,B)$ in any part of the paper. Instead, we have used the Noether’s symmetry equations \eqref{noether-eqs} and we have solved these equations for different branches (Case 1 to Case 4) appearing in those equations. These equations depend on $f(T,B)$ and the Noether’s vector. To find out the form of $f(T,B)$, we solved the system of 34 partial differential equations admitting a non-zero Noether’s vector. Thus, the Noether’s symmetry equations select the form of the function $f(T,B)$ and the form of the Noether’s vector. It is interesting to mention that, for example, a power-law $f(T)=f_0 T^n$ directly appears from these equations, and several $f(T)$ studies have been carried out in this model finding different astrophysical and cosmological features such as the possibility of alleviating the $H_0$ tension in cosmology~\cite{Nunes:2018xbm} or the possibility of explaining galaxy rotation curves in astrophysics~\cite{Finch:2018gkh}. Furthermore, different $f(T,B)$ power-law forms found in this paper have been also used (by hand) in other papers such as \cite{Escamilla-Rivera:2019ulu,Bahamonde:2016cul,Zubair:2018wyy} for explaining cosmological viable models.\\

It is still unclear if the solutions found might be astrophysically interesting. One can follow a similar approach as \cite{Bahamonde:2019zea} and study possible effects such as the photon sphere and perihelion shift of those solutions. Another study that can be done is to analyse the accretion process of these solutions~\cite{Bahamonde:2015uwa,Ahmed:2016cuy}. Still, there is a long route for understanding the possible astrophysical effects of modified Teleparallel theories of gravity. It might be interesting to follow a similar approach done in this paper for other Teleparallel theories, such as scalar-tensor theories~\cite{Bahamonde:2018miw,Bahamonde:2019shr,Bahamonde:2015hza,Hohmann:2018ijr,Hohmann:2018dqh,Hohmann:2018dqh,Hohmann:2018vle,Hohmann:2018rwf,Zubair:2016uhx,DAgostino:2018ngy,Abedi:2018lkr}, non-local theories~\cite{Bahamonde:2017bps} or theories with non-minimally couplings between matter and gravity~\cite{Bahamonde:2017ifa}. These works will be done in forthcoming studies. 

\begin{table}[H]
\centering
\scalebox{1.2}{\bgroup
\def\arraystretch{3} \begin{tabular}{|c |c | c |} 
\hline
 $f(T,B)$ & $a(r)$ & $b(r)$  \\ [0.5ex] \hline 
$f_0 T^{n}$ & $\left( \displaystyle\frac{r}{r_0} \right)^{\frac{4 n (n-1)(2n -3)}{4 n^2 -8 n +5}}$  & $\displaystyle\frac{ (2n-1)(4n -5)}{4 n^2 - 8n +5}$  \\ [1ex] \hline 
$f_0 \displaystyle T^{\frac{ (k+2)(k-1 +q) \pm \tilde{C} }{ 4( k -1 + q) }}+f_1 B^{\frac{ (k+2)(k-1 +q) \pm \tilde{C} }{ 4( k -1 + q) }}$& $\displaystyle\Big(\frac{r}{r_0}\Big)^q$&$k$ \\ [0.5ex] \hline 
 $f_0 T+f_2 B \log B$& $\left( \displaystyle\frac{r}{r_0} \right)^{-\frac{f_2 \left(3 f_0^2+f_0 f_2-2 f_2^2\pm 2 K_1\right)}{f_0 \left(f_0^2+2 f_0 f_2+5 f_2^2\right)}}$ & $\frac{f_0^2 f_2+8 f_0 f_2^2\pm K_1 (f_0+f_2)+f_2^3}{f_0 \left(f_0^2+2 f_0 f_2+5 f_2^2\right)}$  \\ [0.5ex] \hline 
$\tilde{f_0}T+f_2B\log T$ & $\displaystyle\Big(\frac{r}{r_0}\Big)^{q}$ & $ \frac{K_2^2+K_2 (q+3)-11 q^2+18 q+21}{6 K_2}$ \\ [0.5ex] \hline 
 $f_0 B\, T^{\frac{ (q+1)( 3 q^3 + 9 q^2 + 20 q + 16) \pm q^2 C }{4 (q+2) \left(2 q^2 + 5 q + 4 \right)}}$ & $\Big(\displaystyle\frac{r}{r_0}\Big)^q$ & $\frac{ ( q+8)( q+1)^2 \mp q C }{ 2 ( 2 q^2 + 5 q + 4) }$\\ [0.5ex] \hline 
  \end{tabular}\egroup}
 \caption{Exact solutions in modified Teleparallel gravity described by $f(T,B)$ where the metric is $ds^2=a(r)^2dt^2-b(r)^2dr^2-r^2d\Omega^2$. The extra constants appearing in the table are: $K_1^2= f_0^4+2 f_0^3 f_2-3 f_0^2 f_2^2+14 f_0 f_2^3+f_2^4$, $K_2^3=(q+3) (37 q^2+ 24 q -27)  + 6 \sqrt{75 q^6+96 q^5+684 q^4+702 q^3-714 q^2-864 q-75}$, $C^2 = q^4 + 4 q^3 + 22 q^2 + 48 q + 33$, $\tilde{C}^2= 4 q^4 + 4 (2-k) q^3 -3( k^2 -8k +4) q^2 - 2(k-1)(k+2) (k-4) q -(k-1)^2 (7 k^2 -16)$, and $f_1$ and $\tilde{f}_0$ are displayed in Eqs.~\eqref{f1is} and \eqref{f0tildeis} respectively} \label{Table1}
 \end{table}


\section*{Acknowledgements}
S.B. is supported by Mobilitas Pluss No. MOBJD423 by the Estonian government.
\appendix
\section{Noether's symmetry equations }\label{appendix}
 
The Noether symmetry condition \eqref{ngs-eq} applied to the Lagrangian \eqref{L} for the metric \eqref{metric} gives $34$ partial differential equations as follows
\begin{eqnarray}
& & f_T \xi_{,a} = 0, \quad  f_T \xi_{,b} = 0, \quad  f_T \xi_{,_M} = 0, \quad  f_{T} \xi_{,_T} = 0, \quad f_{T} \xi_{,_B} = 0, \quad  f_{_{TB}} \xi_{,a}  = 0, \quad   f_{_{BB}} \xi_{a} = 0, \nonumber \\ & &  f_{_{TB}} \xi_{,b} = 0, \quad f_{_{BB}} \xi_{,b} = 0,  \quad f_{_{TB}} \xi_{,_T}  = 0, \quad f_{_{BB}} \xi_{,_B} = 0, \quad  f_{_{TB}} \xi_{,_M}  = 0, \nonumber \\& &  f_{_{BB}} \xi_{,_M} = 0, \quad  f_{_{BB}} \xi_{,_T}  + f_{_{TB}} \xi_{,_B} = 0, \quad ( f_T + f_B ) ( M \eta^1_{,_b} + a \eta^3_{,_b} ) +  \frac{1}{4} \left( V \xi_{,_b}  + K_{,_b} \right) = 0,   \nonumber \\& &  ( f_T + f_B )\left( \frac{\eta^3}{M}  + \eta^1_{,_a}  + \frac{a}{M} \eta^3_{,_a} \right) +  ( f_{_{TT}} + f_{_{TB}} ) \eta^4  + ( f_{_{TB}} + f_{_{BB}} ) \eta^5  +  \frac{M}{2 b} \left( f_{_{TB}} \eta^4_{,r} + f_{_{BB}} \eta^5_{,r} \right) + \frac{1}{4 M} \left( V \xi_{,_a}  + K_{,_a} \right) = 0,  \nonumber \\& &   ( f_T + f_B ) \left( \frac{\eta^1}{a} + \frac{M}{a} \eta^1_{,_M} + \eta^3_{,_M} \right) +  ( f_{_{TT}} + f_{_{TB}} ) \eta^4 + ( f_{_{TB}}  + f_{_{BB}} ) \eta^5 - \frac{f_T}{a b} \left( M \eta^1_{,r}  + a \eta^3_{,r} \right) \nonumber \\& &   \qquad \qquad + \frac{M}{b} \left(  f_{_{TB}} \eta^4_{,r} + f_{_{BB}} \eta^5_{,r} \right) + \frac{1}{4 a} \left( V \xi_{,_M}  + K_{,_M} \right) = 0,   \nonumber \\& &   ( f_T + f_B  ) \left( M \eta^1_{,_T}  + a \eta^3_{,_T} \right) + \frac{M}{2 b} f_{_{TB}} \left( M \eta^1_{,r} + 2 a \eta^3_{,r}  \right) + \frac{1}{4} \left( V \xi_{,_T}  + K_{,_T} \right) = 0,   \nonumber \\& &  ( f_T + f_B ) ( M \eta^1_{,_B}  + a \eta^3_{,_B} ) + \frac{M}{2 b} f_{_{BB}} ( M \eta^1_{,r}  + 2 a \eta^3_{,r} ) + \frac{1}{4} \left( V \xi_{,_B}  + K_{,_B} \right) = 0,  \nonumber \\& &  2 f_T \eta^3_{,_a}  - M ( f_{_{TB}} \eta^4_{,_a}  + f_{_{BB}} \eta^5_{,_a}  ) = 0,  \quad  2 f_T \eta^3_{,_b}   - M \left( f_{_{TB}} \eta^4_{,_b}  + f_{_{BB}} \eta^5_{,_b}  \right) = 0, \label{noether-eqs} \\& &  f_T \left( - \frac{\eta^2}{b} + \frac{\eta^3}{M} +  \eta^1_{,_a}  + \frac{a}{M} \eta^3_{,_a}  + \eta^3_{,_M} - \xi_{,r}  \right) + f_{_{TT}} \eta^4 + f_{_{TB}} \eta^5  - \frac{M}{2} ( f_{_{TB}} \eta^4_{,_M} + f_{_{BB}} \eta^5_{,_M}  ) - a \left( f_{_{TB}} \eta^4_{,_a}  + f_{_{BB}} \eta^5_{,_a} \right) = 0, \nonumber \\& &    f_{_{TB}} \left( -\frac{\eta^2}{b} + \frac{2 \eta^3}{M}  + \eta^1_{,_a} + \frac{2 a}{M} \eta^3_{,_a} + \eta^4_{,_T} - \xi_{,r}  \right) + f_{_{TBT}} \eta^4 + f_{_{TBB}} \eta^5 - \frac{2 f_T}{M} \eta^3_{,_T} + f_{_{BB}} \eta^5_{,_T} = 0,   \nonumber \\& &   f_{_{BB}} \left( -\frac{\eta^2}{b} + \frac{2 \eta^3}{M}  +  \eta^1_{,_a}  + \frac{2 a}{M} \eta^3_{,_a} + \eta^5_{,_B} - \xi_{,r}  \right) + f_{_{TBB}} \eta^4  + f_{_{BBB}} \eta^5  - \frac{2 f_T}{M} \eta^3_{,_B} + f_{_{TB}} \eta^4_{,_B}  = 0, \nonumber \\& &   f_T ( M \eta^1_{,_b}  + a \eta^3_{,_b} ) - a M ( f_{_{TB}} \eta^4_{,_b}  + f_{_{BB}} \eta^5_{,_b}  ) = 0, \quad  f_{_{TB}} ( M \eta^1_{,_b} + 2 a  \eta^3_{,_b}  ) = 0,  \quad  f_{_{BB}} ( M \eta^1_{,_b} + 2 a \eta^3_{,_b}  ) = 0,  \nonumber \\& &  f_T \left( \frac{\eta^1}{a} - \frac{\eta^2}{b}  + \frac{2 M}{a} \eta^1_{,_M} + 2 \eta^3_{_M} - \xi_{,r} \right) + f_{_{TT}} \eta^4  + f_{_{TB}} \eta^5  - 2 M (f_{_{TB}} \eta^4_{,_M} + f_{_{BB}} \eta^5_{,_M} ) = 0, \nonumber \\& & f_{_{TB}} \left( \frac{\eta^1}{a} - \frac{\eta^2}{b} + \frac{\eta^3}{M} + \frac{M}{2 a} \eta^1_{,_M}  + \eta^3_{,_M}  + \eta^4_{,_T}  - \xi_{,r} \right) + f_{_{TBT}} \eta^4  + f_{_{TBB}} \eta^5 - f_T \left( \frac{1}{a} \eta^1_{,_T}  + \frac{1}{M} \eta3_{,_T} \right) + f_{_{BB}} \eta^5_{,T} = 0, \nonumber \\& &  f_{_{BB}} \left( \frac{\eta^1}{a} - \frac{\eta^2}{b} + \frac{\eta^3}{M} + \frac{M}{2 a} \eta^1_{,_M}  + \eta^3_{,_M} + \eta^5_{,_B}  - \xi_{,r} \right) + f_{_{TBB}} \eta^4  + f_{_{BBB}} \eta^5 - f_T \left( \frac{1}{a} \eta^1_{,_B} + \frac{1}{M} \eta^3_{,_B} \right) + f_{_{TB}} \eta^4_{,_B} = 0,  \nonumber \\& &  f_{_{TB}} ( M \eta^1_{,_T} + 2 a \eta^3_{,_T} ) = 0,  \quad  f_{_{TB}} ( M \eta^1_{,_B} + 2 a \eta^3_{,_B} ) + f_{_{BB}} ( M \eta^1_{,_T} + 2 a \eta^3_{,_T}  ) = 0, \quad   f_{_{BB}} ( M \eta^1_{,_B} + 2 a \eta^3_{,_B} ) = 0,  \nonumber \\& & V_{,_a} \eta^1 + V_{,_b} \eta^2 + V_{,_M} \eta^3  + V_{,_T} \eta^4 + V_{,_B} \eta^5 + 4 ( f_T + f_B ) ( M \eta^1_{,r}  + a \eta^3_{,r} ) + V \xi_{,r}  + K_{,r}  = 0 \nonumber
\end{eqnarray}
where $V \equiv a b \left[ M^2 (T f_{_T} + B f_{_B} - f) - 2 f_{_T} \right]$.

\bibliographystyle{utphys}
\bibliography{references}

\end{document}